\pdfoutput=1

\documentclass[12pt]{article}
\usepackage{graphicx,psfrag,epsf}
\usepackage{enumerate}
\usepackage{natbib}
\usepackage{amsthm, amsmath, amsfonts, mathrsfs, amssymb, bm, latexsym}
\usepackage{mathtools}
\usepackage{graphicx}
\usepackage[T1]{fontenc}
\usepackage{caption}
\usepackage{subcaption}
\usepackage{algorithm}
\usepackage{algpseudocode}
\usepackage{float}
\usepackage{times}
\usepackage{xcolor}
\usepackage{multirow}
\usepackage{authblk}

\newcommand{\blind}{0}
 
\newcommand{\bfx}{{\mathbf{x}}}
\newcommand{\bfX}{{\mathbf{X}}}
\newcommand{\bftheta}{{\mathbf{\theta}}}

\begin{document}

\def\spacingset#1{\renewcommand{\baselinestretch}%
{#1}\small\normalsize} \spacingset{1}


\if0\blind
{
  \title{\bf Model-Free Local Recalibration of Neural Networks}
  \author[1]{R. Torres}
  \author[2]{D. J. Nott}
  \author[3]{S. A. Sisson}
  \author[1]{T. Rodrigues}
  \author[1]{J. G. Reis}
  \author[1]{G. S. Rodrigues}
  \affil[1]{University of Brasília}
  \affil[2]{National University of Singapore}
  \affil[3]{University of New South Wales, Sydney}
  \maketitle
} \fi

\if1\blind
{
  \bigskip
  \bigskip
  \bigskip
  \begin{center}
    {\LARGE\bf Title}
\end{center}
  \medskip
} \fi

\bigskip
\begin{abstract}
Artificial neural networks (ANNs) are highly flexible predictive models.   
However, reliably quantifying uncertainty for their predictions is a continuing challenge. 
There has been much recent work on ``recalibration" of predictive distributions for ANNs, so that forecast probabilities for events of interest are consistent with certain frequency evaluations of them.  Uncalibrated probabilistic forecasts are of limited use for many important decision-making tasks.
To address this issue, we propose a localized recalibration of ANN predictive distributions using the dimension-reduced representation of the input provided by the ANN hidden layers. Our novel method draws inspiration from recalibration techniques used in the literature on approximate Bayesian computation and likelihood-free inference methods. Most existing calibration methods for ANNs can be thought of as calibrating either on the input layer, which is difficult when the input is high-dimensional, or the output layer, which may not be sufficiently flexible.  Through a simulation study, we demonstrate that our method has good performance compared to alternative approaches, and explore the benefits that can be achieved by localizing the calibration based on different layers of the network. 
Finally, we apply our proposed method to a diamond price prediction problem, demonstrating the potential of our approach to improve prediction and uncertainty quantification in real-world applications.
  
\end{abstract}

\noindent%

\spacingset{1.45}
\section{Introduction}
\label{sec:intro}

Artificial neural networks (ANNs) are a highly effective class of predictive models which are widely used in many classification and regression tasks.  
Modern neural networks with higher representation capacity possess the flexibility to capture diverse nonlinear relationships within the modeled system.   
However, improvements in prediction have been associated with poor uncertainty quantification compared to earlier versions of these models 
\citep{guo2017, kuleshov2022, Xiong2023, dheur2023}.  
A possible exception to this phenomenon occurs in
recent state-of-the-art non-convolutional architectures for image classification \citep{minderer2021}.  

Our work is concerned with probabilistic forecasts obtained
from neural networks;
that is, forecasts 
obtained from ANNs which
take the form of probability distributions.  
For a sequence of probabilistic forecasts to
be useful, it is desirable that they should be ``calibrated".  
There are different kinds of calibration, but the
common idea is that the forecast probabilities should
agree with some frequency evaluations of them based
on the observations.  For example, one commonly used type
of calibration would imply that for a sequence of events with
forecast probability $0.5$, half of them should occur in the
limit of an increasing number of forecasts.  
\cite{gneiting2007} suggest that probabilistic forecasting
should aim to achieve sharpness of predictions (low uncertainty)
subject to calibration, and discuss three different
types of calibration -- probabilistic, 
marginal, and exceedance calibration.  
Other notions of calibration have appeared more 
recently in the literature
on uncertainty quantification methods for ANNs (discussed
below).
Calibrated uncertainty quantification is important
in many high stakes applications of neural network models, 
for which quantifying the risk of undesirable outcomes may 
play an important role in decision making.   

Our paper makes three main contributions.  First, we consider
a new approach to the recalibration of ANN predictive
distributions, which localizes the calibration based
on the representation of the input given by the ANN hidden
layers.  Existing methods of recalibration can be thought
of as recalibrating based on the output layer, which
may not be flexible enough, or on the input layer, which
is difficult when the input is high-dimensional.  
The approach we describe is 
inspired by \citet{rodrigues2018}, 
in which the authors calibrate 
approximate posterior distributions in the context of 
likelihood-free inference methods.  
Our second contribution is to give an efficient computational implementation
of this new approach.   We suggest using fast approximate K-Nearest Neighbours (KNN) search algorithms \citep{arya1998} in the
assignment of weights to observations in the recalibration process. 
Our third  contribution is to demonstrate the
good performance of our proposed approach in some real
and simulated examples.   

There is a large existing literature on recalibration
methods for ANNs.  For classification problems, ANNs give
predictive distributions in the form of a vector of probabilities
for a finite number of possible classes for the response.  
In contrast, for regression applications in which the
response is continuous, probabilistic forecasts take the form of a continuous distribution parametrized in some way by the ANN outputs.

Recalibration methods for classification problems have
the more extensive literature, and in this context
\citet{guo2017} observed that increased model capacity and lack of regularization are closely related to miscalibration. To alleviate this problem, they proposed a post-processing technique called temperature scaling, derived from Platt scaling \citep{platt1999}. Temperature scaling involves dividing the network's logit by a scalar value, $T > 0$, before applying a softmax
transformation, which can improve calibration 
without compromising accuracy.
\citet{kumar2019} highlighted that widely used methods such as Platt scaling and temperature scaling may not be as well calibrated as reported. They introduced the scaling-binning calibrator, which bins the values of a parametric fitted function to reduce variance and promote calibration.
Additionally, \citet{mukhoti2020} observed that an increase in the confidence of a network about incorrectly classified test samples is a key indicator of miscalibration. They suggested using focal loss instead of cross-entropy loss to achieve better calibrated models.
More recently, \citet{Xiong2023} argued that existing calibration algorithms commonly neglect the problem of proximity bias, which refers to the tendency of models to exhibit greater overconfidence in low proximity data (i.e.~located in low-density regions of the data distribution) compared to high proximity samples. Their new method, coined \textsc{ProCal}, showed excellent results compared to existing alternative approaches. For a detailed overview of the area of classifier calibration, see \cite{silva2023}.

Less attention has been paid to recalibration methods
in regression problems 
where forecasts are in the form of a probability
distribution for a continuous outcome.  \cite{kuleshov2018}
considered a method for achieving 
quantile calibration, which is a sample version
of probability calibration \citep[Section 3.1]{gneiting2007}, by using isotonic regression 
to learn a transformation for adjusting the quantiles of 
model predictive distributions.  
See also \cite{menendez+fgs14}.
The method of \cite{kuleshov2018} achieves uniformity of probability
integral transform (PIT) values, which have long been
used for forecast evaluation \citep{Dawid1984}.  \cite{utpala2020} suggest achieving calibration implicitly by 
including a penalty term in the training loss to encourage 
uniformity of the PIT values.  Building on this approach, 
\cite{dheur2023} introduce a new 
regularization objective which uses kernel
density estimation (KDE) and an approximation of the indicator 
function which makes the PIT distribution function differentiable.
They also explore links between quantile calibration and 
conformal prediction.

The adjustment approaches discussed above
are global in nature,
and make the implicit assumption that the biases of the model remain consistent across the entire covariate space.
In contrast, \cite{song2019} consider the notion
of distribution calibration.  
It requires equality between the forecast distribution
for some given input, and the true conditional distribution
of the outcome when we consider the input as a random variable
and condition on this random input 
leading to the same forecast distribution
as the given input.  
They propose a method for post-hoc distribution recalibration
which involves learning a Beta calibration map \citep{kull2017a, kull2017b}, which varies 
in the parametrization of the predictive distribution.  The calibration map is learnt using a multioutput Gaussian process model and a complex approximate variational inference scheme to estimate multiple parameters, including kernel parameters, variational parameters, and link parameters.

Recently, \cite{kuleshov2022} proposed a post-hoc recalibration method for regression models which is considered to be at the forefront of current techniques. Their approach involves fitting a quantile function regression model \citep{si2021} taking a low-dimensional feature set derived from the predictive distribution as an input.  Typically this input is based on the output layer of the original network, which is expanded by adding a feature $\tau \in [0, 1]$. The recalibration 
model then provides a direct estimate of the $\tau$-th quantile of the distribution of the response, and parameters
of the recalibration model are learned using a
training criterion which employs a quantile check loss.
The key insight is that while fitting the regression model over the 
parametrization used for the predictive distributions is generally
efficient, it is not as straightforward when applied directly to the
possibly high-dimensional neural network input. Hence, the original model serves to reduce the original input into a set of summary statistics which are subsequently fed into the calibration model.

The approaches built on the definition of distribution calibration are capable of operating locally. However, the notion of localization is viewed with respect to the ANN's output space. This is in contrast to a number of recalibration methods that aim to correct the model's local biases in the actual input space \citep{prangle2014, rodrigues2018, tran2020}. We demonstrate that an intermediate layer (or the input layer) may provide a richer space for diagnosing local biases while still operating on a low-dimensional space that avoids the curse of dimensionality. 

This document is organized as follows. In Section 
\ref{sec:background} we give further background about 
existing definitions of calibration and recalibration methods
based on them.  
Section \ref{sec:method} then presents our new recalibration method. In Sections \ref{sec:gaussian} and \ref{sec:gamma}, we apply the proposed method to a non-linear Gaussian model and a Gamma model, respectively, and discuss its impact on performance. In Section \ref{sec:simulation}, we run a simulation study to evaluate how the proposed method fares in a variety of scenarios, including different recalibration configurations and network models. In Section \ref{sec:diamonds} we test our method with a real data set example. Lastly, we summarise the results in Section \ref{sec:conclusion}.

\section{Background}
\label{sec:background}

We are interested in estimating the conditional distribution of a
continuous target random variable $Y \in \mathcal{Y}$ given 
an input $\bfX \in \mathcal{X}$, $F(Y|\bfX)$, from a training set $\mathcal{D} = \{(\bfx_i, y_i)\}_{i=1}^N$ consisting of $N$ independent identically distributed (i.i.d.) samples from the joint distribution of $(\bfX, Y)$. A probabilistic ANN defines a function $H : \mathcal{X} \to \mathcal{F}$ that maps an input vector $\bfX$ to a predictive cumulative distribution $\hat{F}(Y | \bfX) : \mathcal{Y} \to [0, 1]$ within the space $\mathcal{F}$ of probability distributions over $\mathcal{Y}$. 
This study specifically focuses on this setting.  
It will be assumed that $\hat{F}(Y|\bfX)$ is continuous
and strictly increasing for each $\bfX\in \mathcal{X}$, 
and we write $\hat{F}^{-1}(p|\bfX)$ for its inverse.  
Where convenient we will write $\hat{F}_i(Y)$ 
and $\hat{F}_i^{-1}(p)$ for $\hat{F}(Y|\bfx_i)$ and
$\hat{F}^{-1}(p|\bfx_i)$ respectively.  
We now give some background on the notions
of quantile and distribution calibration.

\noindent
\textbf{Quantile calibration:} \quad \cite{kuleshov2018} suggest the following definition of calibration for which they introduced
the term 
quantile calibration. Given $\mathcal{D}$ as above, a neural network is said to be quantile-calibrated if
\begin{equation} \label{eq1.1}
\frac{\sum_{i = 1}^N \mathbb{I}{\{y_i \le \hat{F}_i^{-1}(p)\}}}{N} \rightarrow p, \space \qquad\forall \: p \in [0, 1],
\end{equation}
as $N \rightarrow \infty$, where $\mathbb{I}\{A\}$ denotes the indicator function which equals 1 if $A$ is true and 0 otherwise. 
\cite[Theorem 2]{gneiting2007} consider this as a sample
version of `probability calibration', and establish an equivalence
between quantile and probability calibration for 
`*-mixing' sequences of random variables \citep{Blum1963}, where
the convergence in \eqref{eq1.1} is almost convergence.  
Probability calibration
requries that $N^{-1}\sum_{i=1}^N F_i(\hat{F}_i^{-1}(p))$ converges
to $p$ for every $p\in [0,1]$ almost surely.

Quantile calibration guarantees that observations fall below the $p$-quantile of the forecast distributions with relative frequency $p$.  Sections \ref{sec:gaussian} and \ref{sec:gamma} demonstrate that this notion of calibration is too weak for
some purposes;  for the models considered there, although approximately 95\% of the test samples fall within 95\% prediction intervals after calibration, this calibration
does not hold after selecting data according
to features of the predictive distribution or the input.

Under the assumption of a continuous and strictly
increasing predictive distribution function, 
quantile calibration is 
equivalent to requiring uniformity of the limiting empirical
distribution of probability 
integral transform (PIT) values $p_i=\hat{F}_i(y_i)$.  
Empirical comparison 
of the PIT values to a uniform distribution 
has been widely used for forecast evaluation \citep{Dawid1984}.
If forecasts are ideal in the sense that the forecast
distribution always coincides with the distribution of the observations, 
then $p_i \sim \text{Uniform}(0, 1)$ for all $i$.   
Formal tests for uniformity of the PIT values and 
descriptive measures of their departure from uniformity based on
the Wasserstein distance and the Cramér-von Mises distance have been
suggested \citep{dheur2023}.

\citet{kuleshov2018} proposed learning a distribution function $\hat{R}:[0, 1] \to [0, 1]$ 
that estimates the CDF of the PITs, $R$, so that forecasts $\tilde{F} = \hat{R} \circ \hat{F}$ are nearly calibrated; if $\hat{R} = R$, then uniformity of the PIT values follows, since
\begin{align*}
\text{Pr}(\tilde{F}_i(y_i) \leq \alpha) 
= \text{Pr}(\hat{F}_i(y_i) \leq \hat{R}^{-1}(\alpha))
= R(\hat{R}^{-1}(\alpha))
= \alpha.
\end{align*}
The authors suggest approximating $R$ by fitting an isotonic regression over a \emph{recalibration set} 
\begin{equation*} \label{eq:kuleshov}
\mathcal{D}_{\text{rec}} = \{ (p_{\text{rec}}^{(i)}, \hat{p}_{\text{rec}}^{(i)}) \}_{i=1}^n,
\end{equation*}
where $p_{\text{rec}}^{(i)} = \hat{F}_i (y^{(i)}_{\text{rec}} | \bfx^{(i)}_{\text{rec}})$ 
is the PIT value computed over the sample $(y^{(i)}_{\text{rec}}, \bfx^{(i)}_{\text{rec}})$, and
\begin{equation*} \label{eq:kuleshov}
\hat{p}_{\text{rec}}^{(i)} = \frac{1}{M} \sum_{j=1}^n \mathbb{I}{\{p_{\text{rec}}^{(j)} \leq p_{\text{rec}}^{(i)}\}}
\end{equation*}
denotes an empirical estimate of $R(p_{\text{rec}}^{(i)})$. 
The recalibration data may not correspond to the training data $\mathcal{D}$.

\noindent
\textbf{Distribution calibration:} \quad A more strict definition of calibration is introduced in \cite{song2019}. Suppose that the
random vector 
$(Y,\bfX)=(Y(\omega),\bfX(\omega))$ is defined on a probability space
$(\Omega,\mathcal{A},P)$.  Let $A(\bfx)=\{\omega\in \Omega: 
\bfX(\omega)=\bfx', \hat{F}(y|\bfx)=\hat{F}(y|\bfx'), 
\forall y\in \mathcal{Y}\}$  
and denote the distribution function of the random variable
$Y|A(\bfx)$ as $F(Y|A(\bfx))$.  
Then the collection of distributions
$\{\hat{F}(Y|\bfx), \bfx\in \mathcal{X}\}$ are distribution
calibrated if
\begin{equation} \label{eq:Song}
F(y | A(\bfx)) = \hat{F}(y| \bfx),\;\;\mbox{for every $y\in \mathcal{Y}$, $\bfx\in\mathcal{X}$}.
\end{equation}

This definition requires that when considering the input $\bfX$
as a random variable, when conditioning on obtaining the
same predictions as the fixed input $\bfx$, the distribution of
the outcome is equal to the forecast distribution for $\bfx$.
This gives a notion of calibration that is local in the output
space.  Distribution calibration implies quantile calibration 
but the reverse implication does not hold. If $\hat{F}(y|\bfx)$ is a parametric distribution indexed by $\bftheta(\bfx)$, which
we write as $\hat{F}(y|\bftheta(\bfx))$, then 
often the values of $\bftheta(\bfx)$ are given
by the components of the neural network's output layer, 
and the condition $\hat{F}(y|\bfx)=\hat{F}(y|\bfx')$ in the
definition of $A(\bfx)$ reduces to $\bftheta(\bfx)=\bftheta(\bfx')$.

\section{Methods}
\label{sec:method}

Inspired by this definition, we 
introduce a novel algorithm that attempts to achieve
uniformity of the PIT values locally.   
Algorithm \ref{alg:alg1} introduces our new post-processing method for recalibrating the predictive distributions of a probabilistic ANN regression model. We assume an $L$-layers ANN has been previously fitted, with the predictive distribution defined according to the network's loss function. For example, if the Mean Squared Error (MSE) was adopted as the loss function, the neural network provides Maximum Likelihood Estimates (MLE) for the mean of a conditional homoscedastic Gaussian distribution. Other approaches, including the non-parametric use of \textit{dropout layers} for estimating the predictive distributions, are discussed in Section \ref{sec:gamma}. 

\begin{algorithm}
\caption{Model-Free Local Recalibration of Neural Networks}
\begin{algorithmic}[1]
\vspace{.25cm}
\State \textbf{Input} \\
\begin{itemize}
\item Recalibration set, $\{y^{(i)}_{\text{rec}}, \bfx^{(i)}_{\text{rec}}\}_{i=1}^{n}$, and new set, $\{\bfx^{(j)}_{\text{new}}\}_{j=1}^{m}$.
\item A neural network and its associated predictive distribution, $\hat{F}(\cdot \,| \, \bfX)$.
\item A positive integer $l$ defining the network's layer where the samples are to be compared.
\item Neural network's outputs of the $l$-th layer on the recalibration set, $\{\mathbf{h}^{(i)}_{\text{rec}}\}_{i=1}^{n}$.
\item A smoothing kernel $K_u(d)$ with scale parameter $u>0$, which may be defined indirectly from a positive integer $k$ that represents the number of observations to be used for recalibration.
\end{itemize}
\vspace{.25cm}

\noindent
\textbf{Cumulative probabilities (PIT-values)}   \vspace{.05cm}
\For{$i \gets 1$ \textbf{to} $n$}
    \State \label{step:pvalues} Set $p_{\text{rec}}^{(i)} = \hat{F}_i (y^{(i)}_{\text{rec}} | \bfx^{(i)}_{\text{rec}})$.
  \EndFor  \vspace{.25cm}

\noindent
\textbf{Recalibration}   \vspace{.05cm}
\For{$j \gets 1$ \textbf{to} $m$}
    \State Compute $\mathbf{h}^{(j)}_{\text{new}}=g(\bfx^{(j)}_{\text{new}})$, where $g$ denotes the network's mapping to the $l$-th layer.
    \State \label{step:distances} Apply the approximate KNN search method to identify the set of indices, $I_j$, corresponding to the observations in $\{y^{(i)}_{\text{rec}}, \bfx^{(i)}_{\text{rec}}\}_{i=1}^{n}$ for which $\|\mathbf{h}^{(i)}_{\text{rec}}-\mathbf{h}^{(j)}_{\text{new}}\|$ are within the $k$-smallest values.
    \For{$i \in I_j$}
    \State \label{step:ys} Set $\tilde{y}^{(j)}_{i} = \hat{F}^{-1}_j (p_{\text{rec}}^{(i)}|\bfx^{(j)}_{\text{new}})$ and assign it a weight $w_i^{(j)} \propto K_u(\|\mathbf{h}^{(i)}_{\text{rec}}-\mathbf{h}^{(j)}_{\text{new}}\|)$.
    \EndFor
    
  \EndFor  \vspace{.25cm}

\noindent
\textbf{Output}   \vspace{.05cm}

\State A set of $k$ weighted samples $\{(\tilde{y}^{(j)}_{i}, w_i^{(j)})\}$ from the recalibrated predictive distribution $$\tilde{F}_j (\cdot \,| \, \bfx^{(j)}_{\text{new}}),$$ for $j = 1, \ldots, m$.

\end{algorithmic}
\label{alg:alg1}
\end{algorithm}

Our recalibration method is composed of two stages: calculating the predictive cumulative probability,  $p_{\text{rec}}^{(i)} = \hat{F}_i (y^{(i)}_{\text{rec}} | \bfx^{(i)}_{\text{rec}})$, for each sample in the recalibration set, $\{y^{(i)}_{\text{rec}}, \bfx^{(i)}_{\text{rec}}\}_{i=1}^n$, and then performing the recalibration for each observation in the \textit{new} set, $\{\bfx^{(j)}_{\text{new}}\}_{j=1}^m$ (a collection of samples for which the inputs are known and for which we want to predict the respective responses).  Throughout this work, we use the validation set (used for tuning the hyperparameters of the neural network) as the recalibration set. 

For each new input $\bfx^{(j)}_{\text{new}}$, we 
compute $\mathbf{h}^{(j)}_{\text{new}}=g(\bfx^{(j)}_{\text{new}})$, where $g$ denotes the network's mapping to the $l$-th layer. Here, $\mathbf{h}^{(j)}_{\text{new}}$ is ideally a nearly sufficient low-dimensional representation (summary statistic) of the raw data implicitly learned by the network during training.
We then use the approximate KNN search method \citep{chen2021} to identify the indices, $I_j$, corresponding to the observations in the recalibration set for which the distances $\|\mathbf{h}^{(i)}_{\text{rec}}-\mathbf{h}^{(j)}_{\text{new}}\|$ are within the $k$-smallest values. 
Lastly, for each $i \in I_j$, we set $\tilde{y}^{(j)}_{i} = \hat{F}^{-1}_j (p_{\text{rec}}^{(i)}|\bfx^{(j)}_{\text{new}})$ and assign it a weight $w_i^{(j)} \propto K_u(\|\mathbf{h}^{(i)}_{\text{rec}}-\mathbf{h}^{(j)}_{\text{new}}\|)$.

Larger weights are therefore given to samples $\tilde{y}^{(j)}_{i}$ for which, according to the metric $\|.\|$, $\mathbf{h}^{(i)}_{\text{rec}}$ and $\mathbf{h}^{(j)}_{\text{new}}$ are similar. This reflects the principle that closer observations are usually more informative of the model's predictive capabilities (at that particular location of the space induced by the network's $l$-layer). The number of observations considered in the recalibration, $k$, plays a critical role in the recalibration performance -- the smaller it is, the better the local bias is captured, at the cost of an increased approximation error associated with the smaller number of weighted samples from the recalibrated distribution. If one wishes to recalibrate a model globally, without considering local biases, it can be achieved by setting $k=n$ and choosing a constant weighting function 
in Step \ref{step:ys} of Algorithm \ref{alg:alg1}. 

The task of identifying the $k$ nearest neighbours of $\mathbf{h}^{(j)}_{new}$ in the recalibration set (Step \ref{step:distances}, Algorithm \ref{alg:alg1}) is very important for algorithm efficiency. 
Due to the curse of dimensionality, it becomes increasingly difficult to find an efficient solution to this problem as the sample size and dimensionality of $\bfX$ increases. Given a set $\mathbf{S}$ of $n$ points in $M$, a $d$-dimensional space, and $q \in M$ a single query point, a naive approach to the problem of finding the closest point in $\mathbf{S}$ to $q$ (through linear search) takes $O(dn)$ time. It is possible to use a GPU for calculating the distances to speed up the process. Trying to find the $k$-closest points of $\mathbf{S}$ to query set $\mathbf{Q}$ is much more costly. An efficient alternative for large data sets is 
approximate KNN search methods. \cite{arya1998}, shows that, given $\epsilon > 0$, the $k$ $(1 + \epsilon)$-approximate nearest neighbors of $q$ can be computed in $O(kd \, \log n)$ time. Recent work on identifying approximate nearest neighbors
\cite{chen2021} can compute a solution with billion-scale data sets twice as fast as previous state-of-the-art search algorithms. 
We explore the influence of the $\epsilon$ approximation parameter in recalibration, along with other parameters, in Section \ref{sec:simulation}.

Let $\mathbf{p} = (p_{\text{rec}}^{(1)}, p_{\text{rec}}^{(2)}, \dots, p_{\text{rec}}^{(n)})$ be a vector of cumulative probabilities, as defined in Step \ref{step:pvalues}. The $j$-th recalibrated predicted value, $\hat{y}_j$, can then be obtained by taking the weighted average of the recalibrated predictive distribution's samples, 
\begin{equation} \label{eq:recal-prediction}
    \hat{y}_j =  \sum_{i \in I_j}  \left(\frac{w_i^{(j)}}{\sum_{i \in I_j} w_i^{(j)}}\right) \tilde{y}^{(j)}_{i},
\end{equation}
where $\tilde{y}^{(j)}_{i} = \hat{F}^{-1}_j (p_{\text{rec}}^{(i)}|\bfx_{\text{new}}^{(j)})$ (see Step \ref{step:ys}).
Interval predictions and estimates of other features of the recalibrated predictive distributions (e.g.~variance) can be similarly calculated. 

A major advantage of the proposed method is that it can be used in any layer of a neural network.
Let $l = 1, \ldots, L$ denote the index of the network layers; with $l = 1$ and $l=L$ indicating that the recalibration is performed in the input and output layers, respectively.
As previously discussed,
current state-of-the-art methods, such as \cite{song2019} and \cite{kuleshov2022}, implicitly recalibrate on $l = L$. In Section \ref{sec:simulation} we 
investigate the effect of calibrating on different layers.  

If performed on the input space ($l=1$) and $\bfX$ is high-dimensional, recalibration can be statistically and computationally
challenging. At the other end of the spectrum, recalibrating on the output layer does not guarantee that the selected observations are, in fact, \textit{close}, since only features of the predictive
distribution are taken into account: two samples far apart in the input space can lead to similar predictions. Recalibrating on an intermediate layer may, therefore, provide two related benefits.  First, the dimensionality can be easily controlled when setting the ANN's architecture. Second, the distances are evaluated over a convenient representation of the original data. 

\section{Illustrations}
\label{sec:verify}

\subsection{Heteroscedastic Gaussian Model}
\label{sec:gaussian}

\begin{figure}
  \centering
  \begin{subfigure}[b]{0.25\linewidth}
    \includegraphics[width=\linewidth]{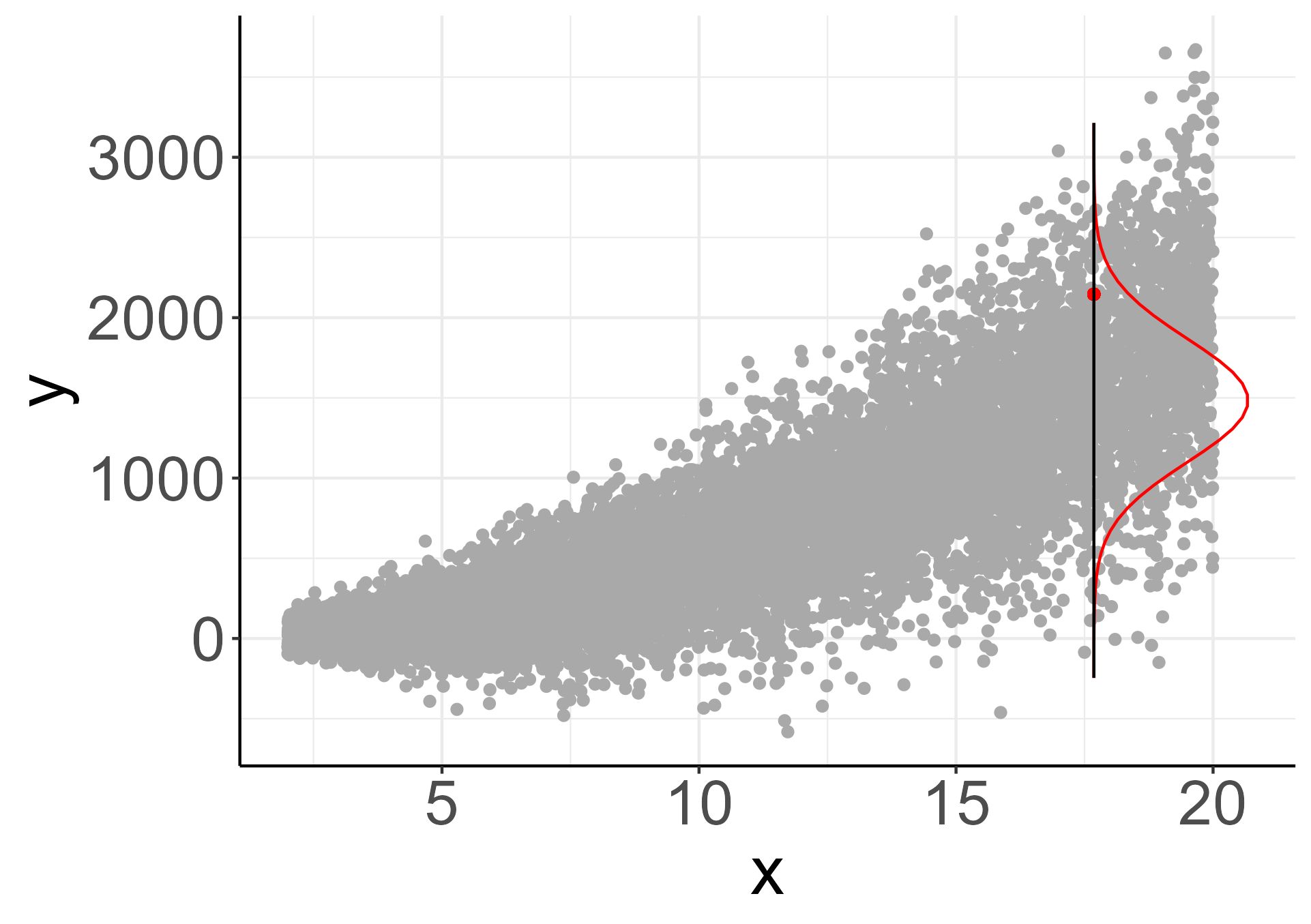}
    \caption{\footnotesize{Scatterplot of Y given X.}}
    \label{fig::3.1a}
  \end{subfigure}
  \vspace{.5cm}
  \begin{subfigure}[b]{0.25\linewidth}
    \includegraphics[width=\linewidth]{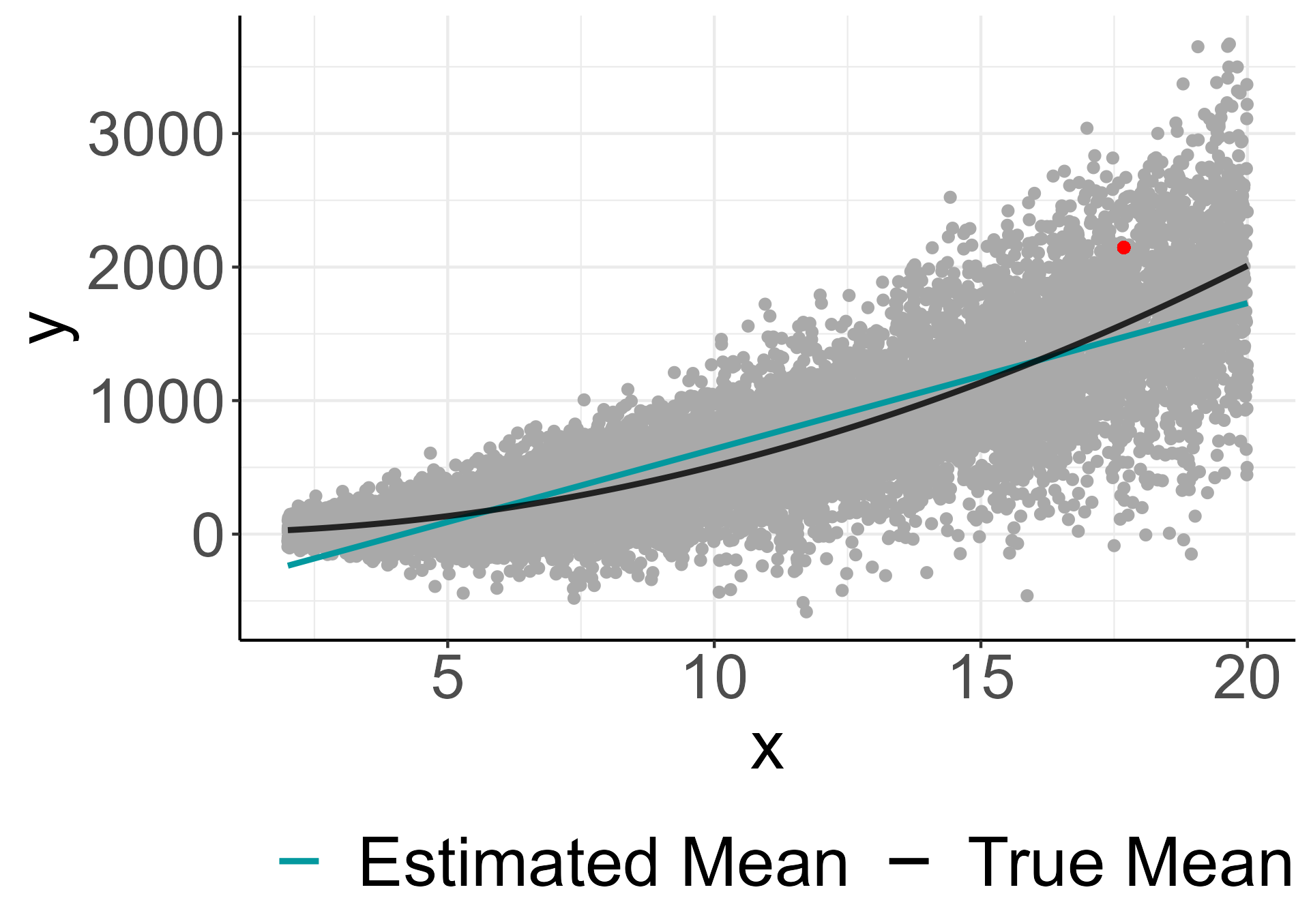}
    \caption{\footnotesize{Mean functions.}}
    \label{fig::3.1b}
  \end{subfigure}
  \vspace{.5cm}
  \begin{subfigure}[b]{0.25\linewidth}
    \includegraphics[width=\linewidth]{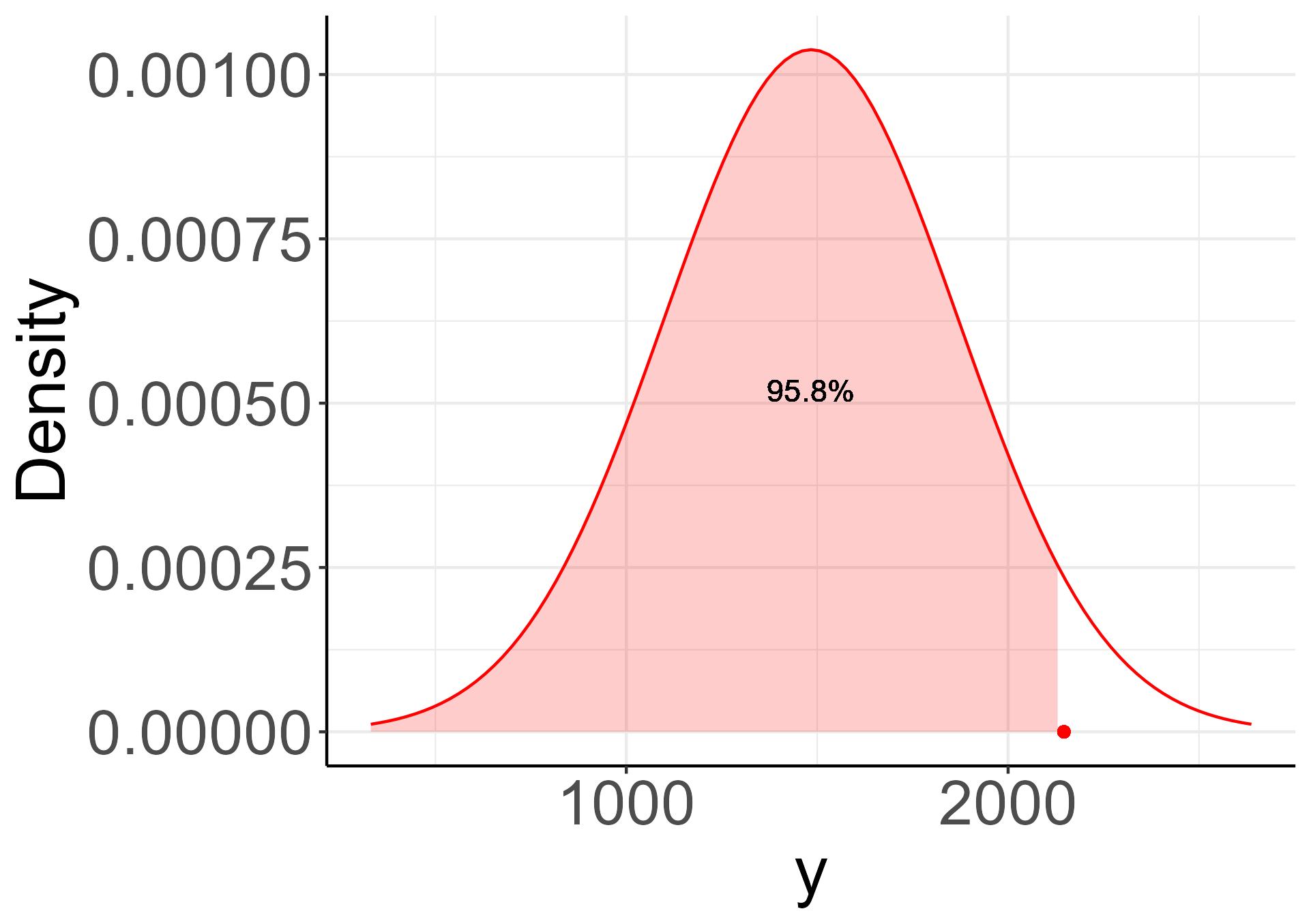}
    \caption{\footnotesize{Predictive density.}}
    \label{fig::3.1c}
  \end{subfigure}
  \vspace{.5cm}
  
  \begin{subfigure}[b]{0.25\linewidth}
    \includegraphics[width=\linewidth]{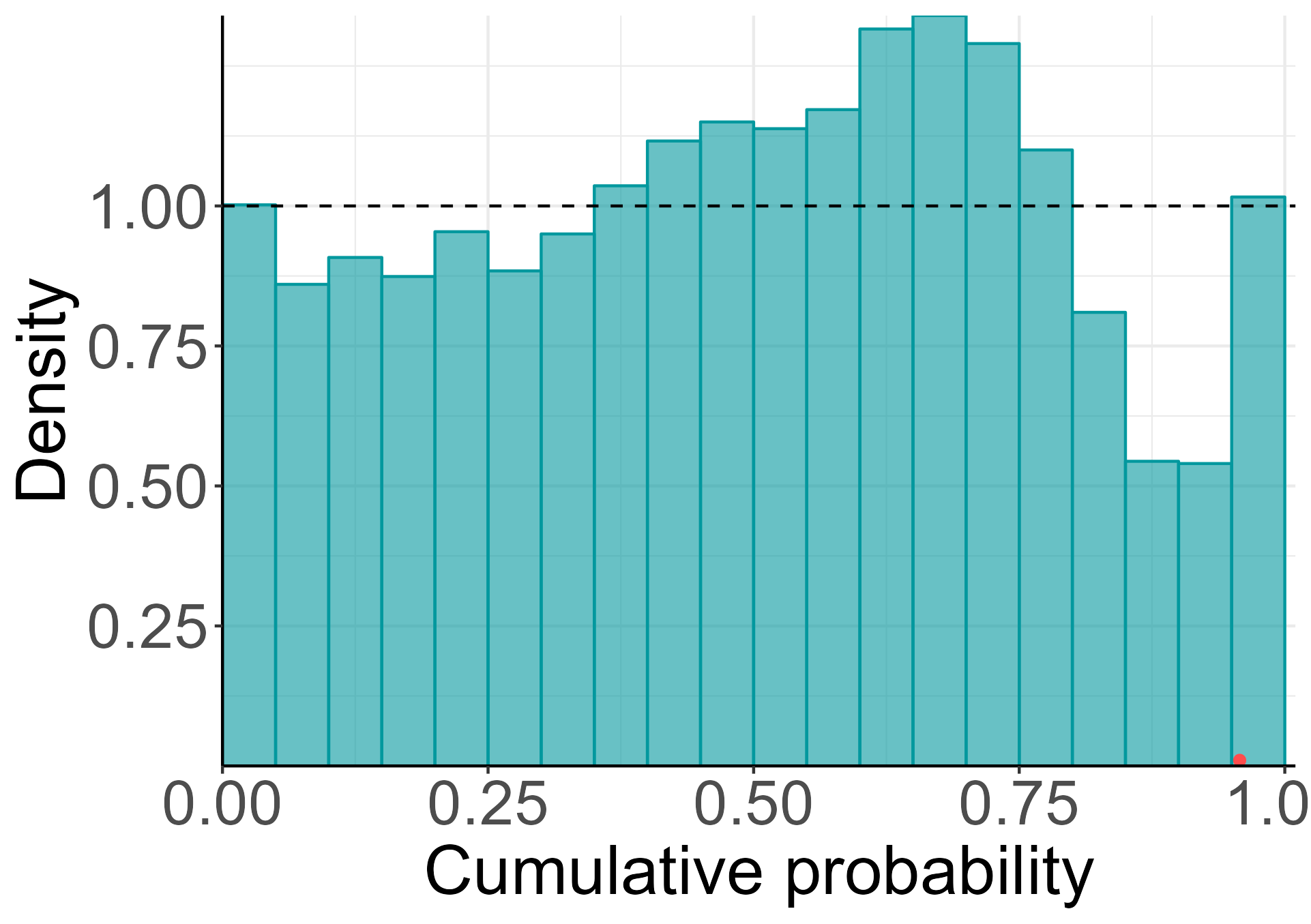}
    \caption{\footnotesize{Global histogram.}}
    \label{fig::3.1d}
  \end{subfigure}
  \vspace{.5cm}
  \begin{subfigure}[b]{0.25\linewidth}
    \includegraphics[width=\linewidth]{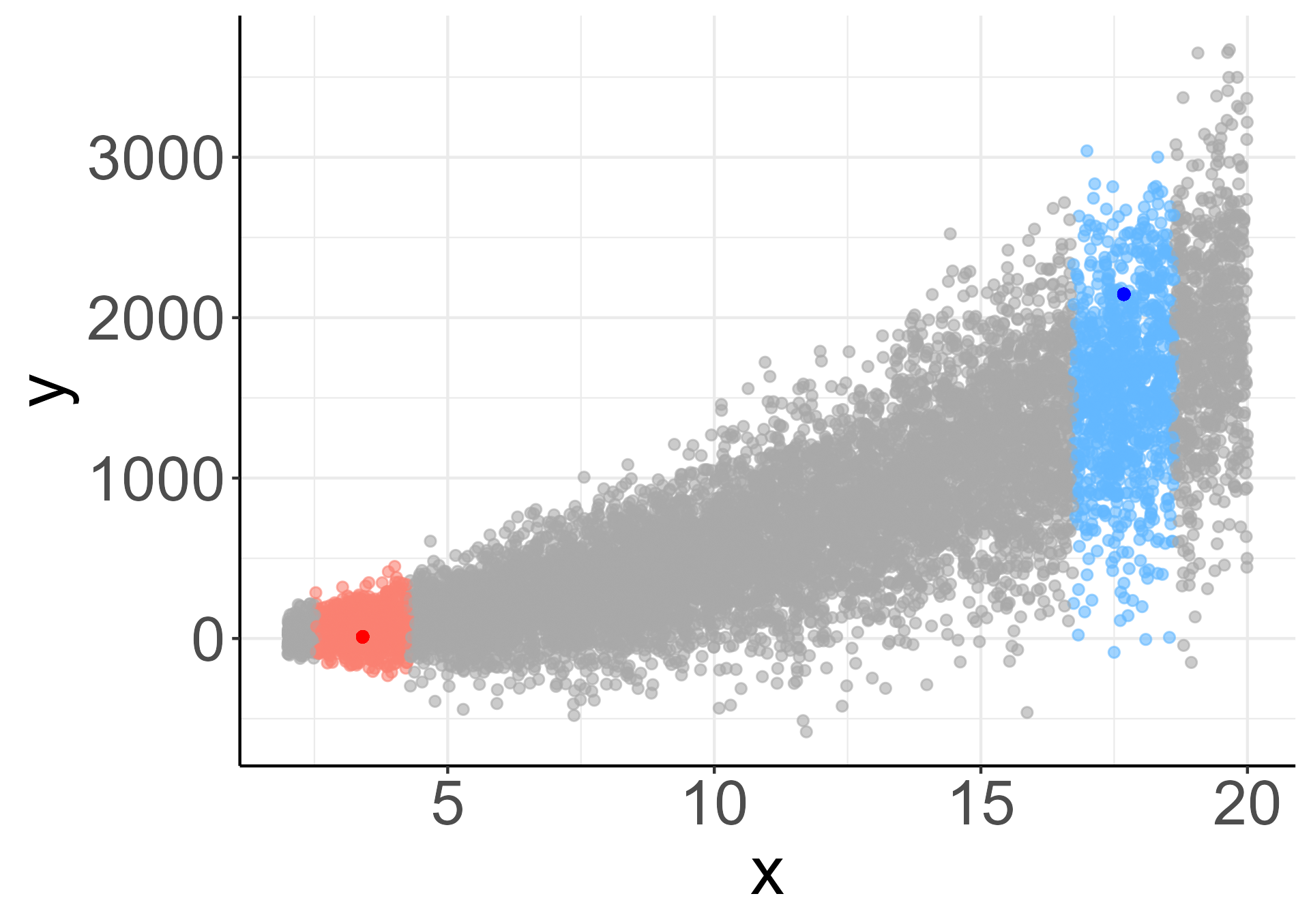}
    \caption{\footnotesize{Distinct neighborhoods.}}
    \label{fig::3.1e}
  \end{subfigure}
  \begin{subfigure}[b]{0.25\linewidth}
    \includegraphics[width=\linewidth]{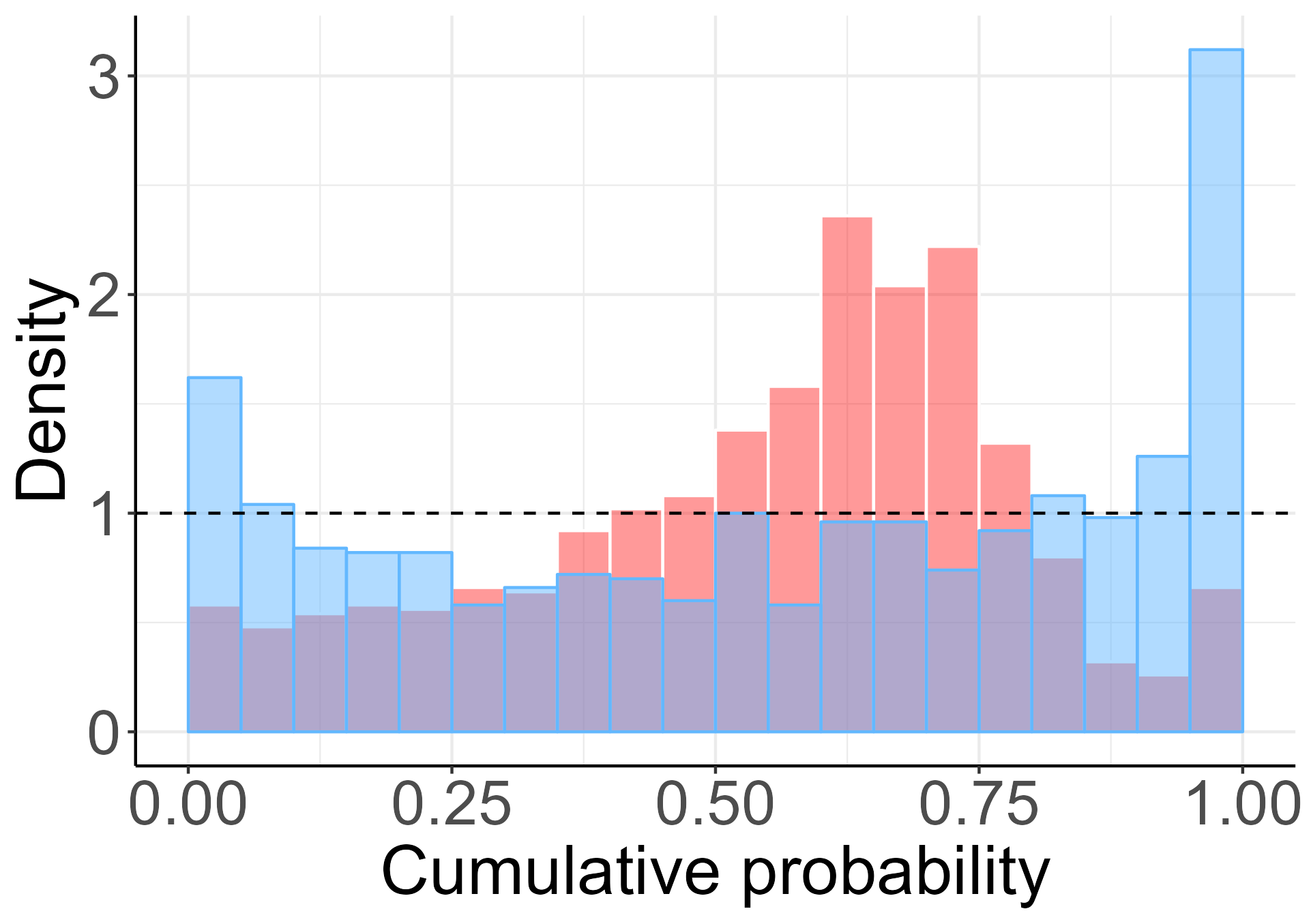}
    \caption{\footnotesize{Local histograms.}}
    \label{fig::3.1f}
  \end{subfigure}
  \caption{\small (a) The true relationship between the response variable and the independent variable, with the position of the highlighted observed point in relation to the predictive distribution given by the model. (b) The true mean contrasted with the model's estimated mean. (c) The predictive density of the highlighted point and its cumulative probability (PIT value). (d) The global cumulative probabilities (PIT) histogram shows the model's global bias. (e) The neighborhood of the two highlighted points. (f) The cumulative probabilities histogram of the model in the two highlighted neighborhoods shows that the model presents two distinct bias patterns depending on the region.}
  \label{fig::3.1}
\end{figure}

Consider $n = 100,000$ samples from the following Gaussian heteroscedastic quadratic model
\begin{equation} \label{eq:sq-gaussian}
Y = 10 + 5X^2 + e,
\end{equation}
where $e \sim N(\mu = 0, \sigma = 30X)$ (Figure \ref{fig::3.1a}) and $\bfX \sim U(2, 20)$. In this study, the samples are split into training ($80\%$), validation ($10\%$) and test sets ($10\%$). 
The validation set is also adopted as the recalibration set. 
We fit the misspecified linear homoscedastic model $\hat{\mu_Y} = \beta_0 + \beta_1X$ over the training data. Figure \ref{fig::3.1b} shows the fitted linear regression line in contrast with the actual mean curve. The red curve shows the plug-in normal predictive distribution 
for the observation $(x,y)=(17.68,2146.22)$ 
which has mean and standard deviation $\hat{y} = 1484.01$ and $\hat{\sigma} = 384.41$, respectively. Figure \ref{fig::3.1c} shows that, for a random variable with distribution $N(\mu = 1484.01, \sigma = 384.41)$, the highlighted observation has an associated cumulative probability of $p = 0.9575$. The non-uniform cumulative probability histogram of all observations in Figure \ref{fig::3.1d} hints at the global bias of the predictions of this model.

Figure \ref{fig::3.1} shows the poor quality of the fit of the linear model due to its misspecification. Frosini's test rejected the null hypothesis of uniformly distributed $p_{\text{rec}}^{(i)}$ for all $i$ with a $0.1\%$ significance level. In the histogram of global cumulative probability estimates (Figure \ref{fig::3.1d}), it is possible to see a combination of bias patterns indicating both model's variance underestimation and overestimation in distinct regions of the covariate space (Figures \ref{fig::3.1e} and \ref{fig::3.1f}). 

We generate samples from the recalibrated distribution $\hat{F}^{-1}_j (\cdot|\bfx^{(j)}_{\text{new}})$ using Algorithm \ref{alg:alg1}, with $l=1$ (that is, we recalibrate in the input space), using Euclidean distance for $\|.\|$, and the Epanechnikov smoothing kernel 
with scale parameter chosen so that $k=1,000$.

\begin{figure}
  \centering
  \begin{subfigure}[b]{0.24\linewidth}
    \includegraphics[width=\linewidth]{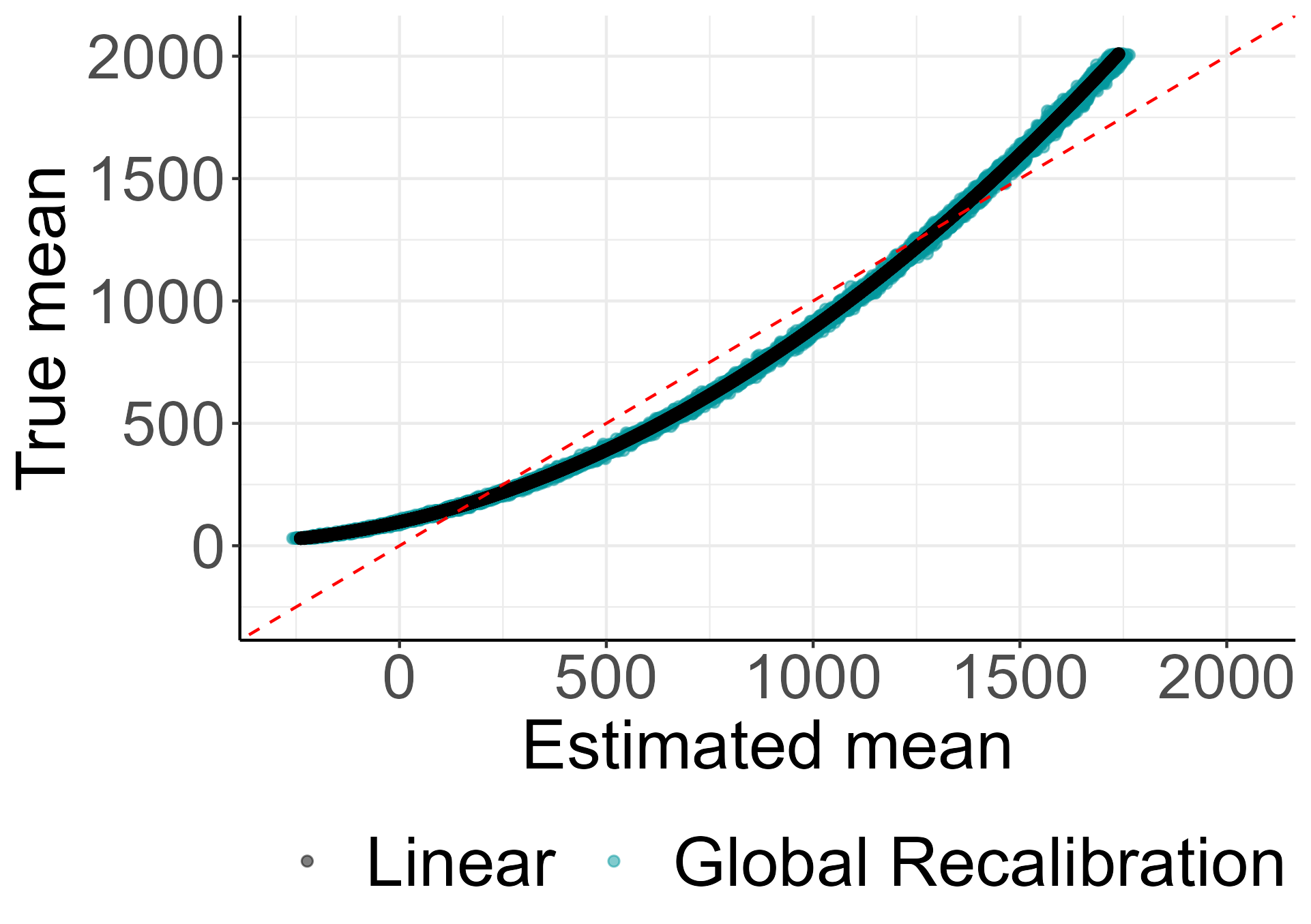}
    \subcaption{\footnotesize{Mean (linear, global).}}
    \label{fig::3.2a}
  \end{subfigure}
  \begin{subfigure}[b]{0.24\linewidth}
    \includegraphics[width=\linewidth]{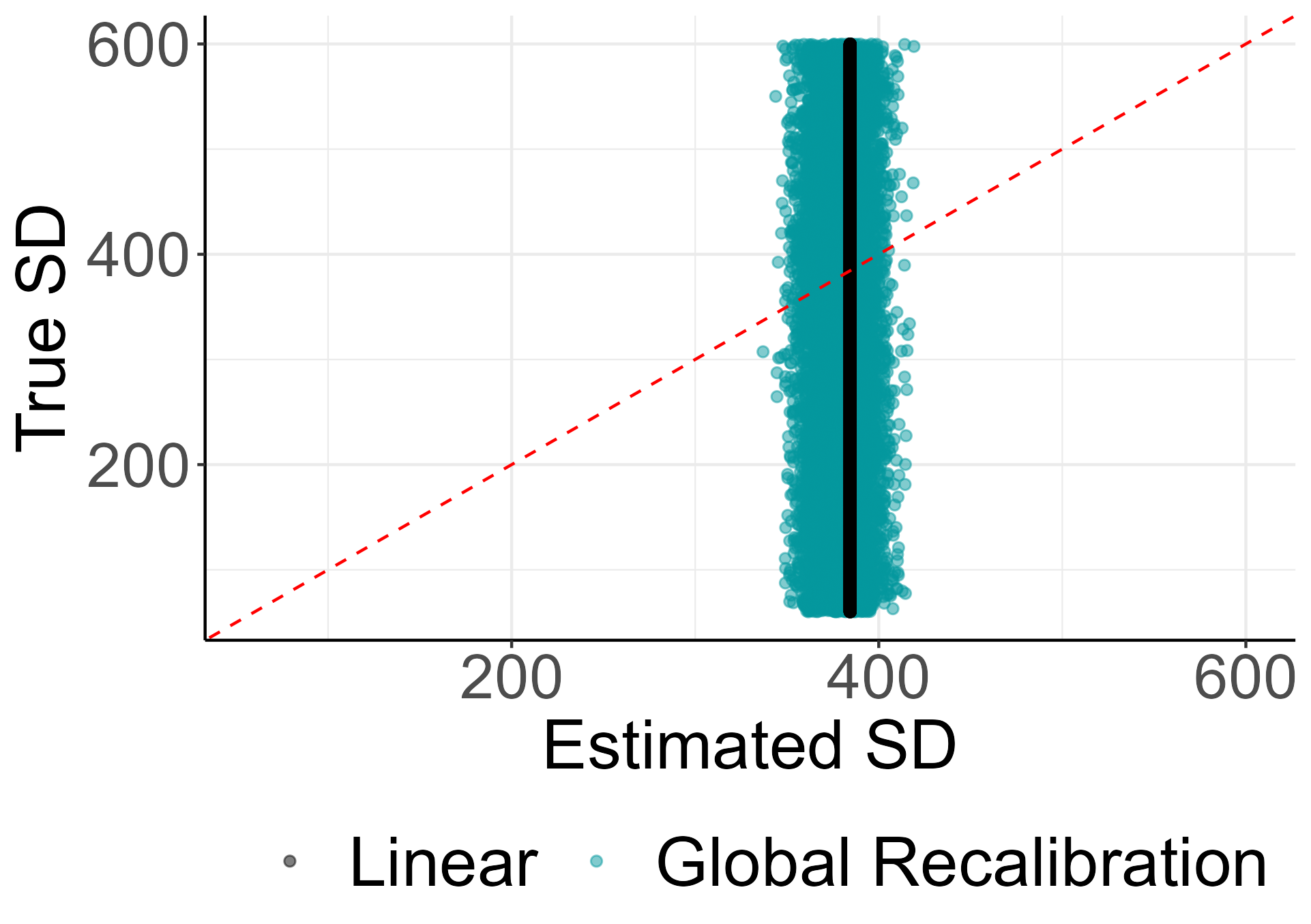}
    \subcaption{\footnotesize{SD (linear, global).}}
    \label{fig::3.2b}
  \end{subfigure} 
  \begin{subfigure}[b]{0.24\linewidth}
    \includegraphics[width=\linewidth]{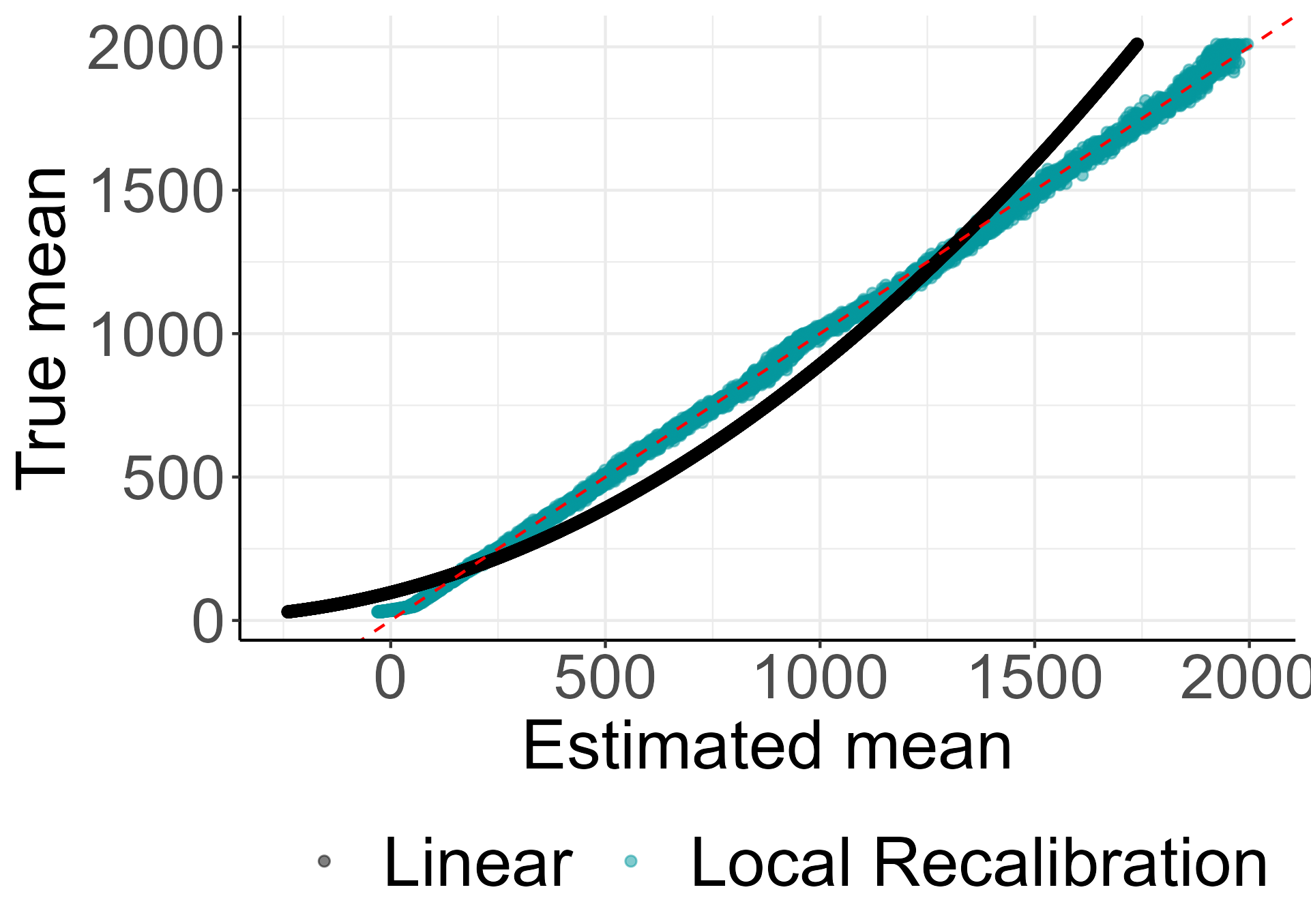}
    \caption{\footnotesize{Mean (linear, local).}}
    \label{fig::3.2c}
  \end{subfigure}
  \begin{subfigure}[b]{0.24\linewidth}
    \includegraphics[width=\linewidth]{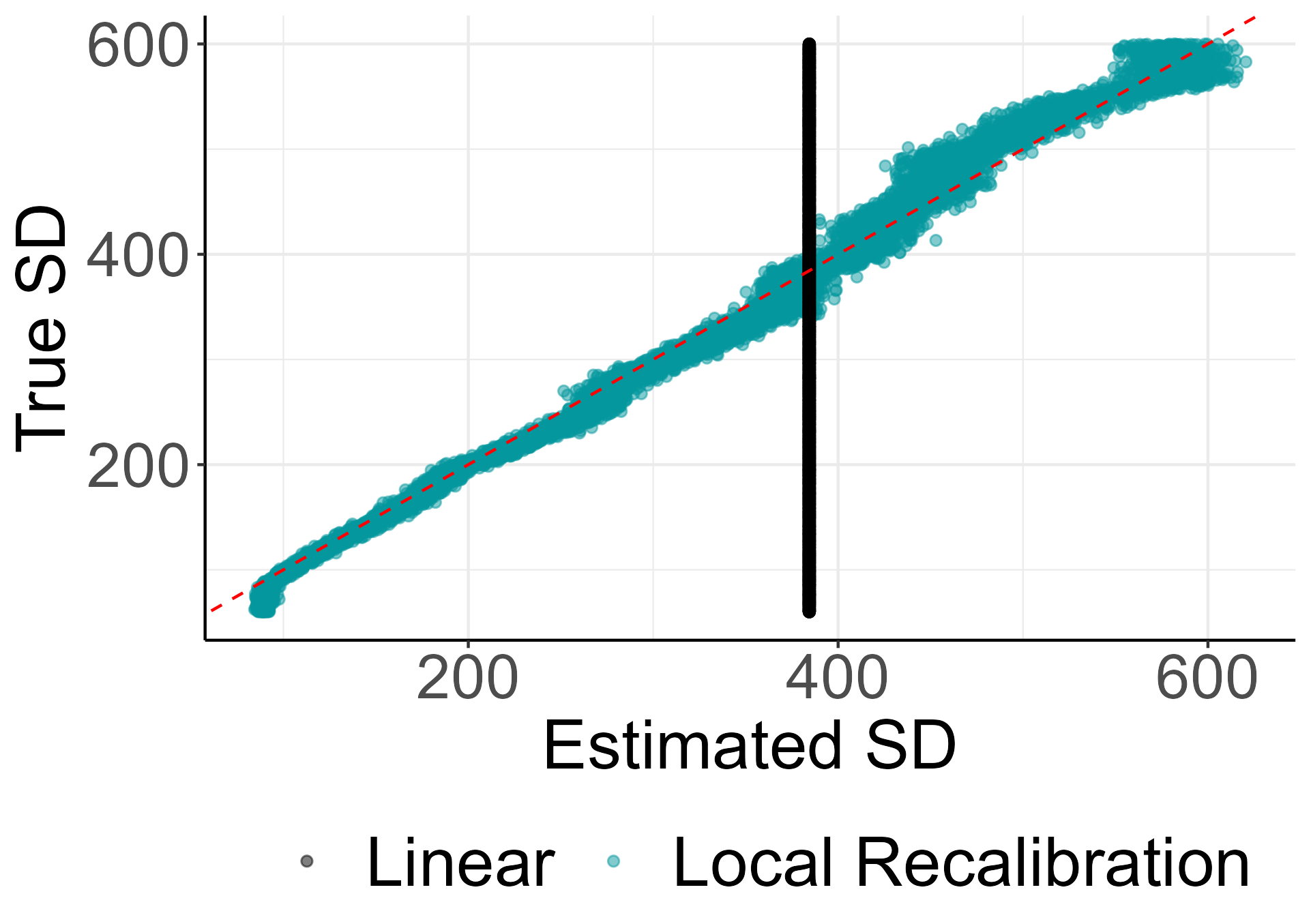}
    \caption{\footnotesize{SD (linear, local).}}
    \label{fig::3.2d}
  \end{subfigure}
  \caption{\small
  In Panel (a), the true and estimated means for each observation in the test set are displayed, both by the linear and the globally-recalibrated model. Panel (b) presents a similar comparison but focuses on the standard deviation. Panels (c) and (d), contrast the linear model with the locally-recalibrated model.}
  \label{fig::3.2}
\end{figure}

Figure \ref{fig::3.2} compares the effect of global and local recalibrations, in terms of how well each recalibrated model estimates the true means and standard deviations. Overall, the global recalibration offered no significant improvement over the linear model, as seen in Figures \ref{fig::3.2a} and \ref{fig::3.2b}. However, the locally-recalibrated model improved both the mean (Figure \ref{fig::3.2c}) and the variance estimations (Figure \ref{fig::3.2d}). 

\begin{table}
\footnotesize
\centering
\begin{tabular}{lcccc}
\hline
Model & MSE &  KL Divergence & Coverage (\%) & sMIS\\
\hline
Linear                         &  14546.25 & 0.8455 & 93.78 & 2.7475 \\
Globally Recalibrated Linear   &  14763.29 & 0.7353 & 94.65 & 2.7391 \\
Locally Recalibrated Linear    &    \textbf{303.93} & \textbf{0.1097} & \textbf{94.96} & \textbf{2.0947} \\
\hline
\end{tabular}
\caption{\small Performance comparison between models via Mean Squared Error (MSE), KL divergence, realised coverage of a 95\% confidence interval, and the standard mean interval score \eqref{ic-score}.} 
\label{table::3.4}
\end{table}

Table \ref{table::3.4} shows performance indicators for all the models considered. The observed coverage represents the percentage of $95\%$ confidence intervals that captured the respective observed value in the test data set. The standard Mean Interval Score (sMIS) \citep{gneiting+r2007} is given by
\begin{equation}\label{ic-score}
IS_{\alpha}(i) = -\bigg((q_{\frac{\alpha}{2}} - q_{1 - \frac{\alpha}{2}}) + \frac{2}{\alpha}(q_{\frac{\alpha}{2}} - y_i)\mathbb{I}_{\{y < q_{\frac{\alpha}{2}}\}}(i) + \frac{2}{\alpha}(y_i - q_{1 - \frac{\alpha}{2}})\mathbb{I}_{\{y > q_{1 - \frac{\alpha}{2}}\}}(i)\bigg),
\end{equation}
where $i = 1, \dots, n$, $q_{\frac{\alpha}{2}}$ and $q_{1 - \frac{\alpha}{2}}$ are the upper and lower interval quantiles, standardized by the validation set mean absolute values. sMIS directly compares prediction intervals by penalizing observations missed and rewarding narrower intervals. It can be seen that local recalibration performed best in all measures compared to the other models.

\subsection{Non-Linear Gamma Model}
\label{sec:gamma}

Here we utilize synthetic data to estimate the Rosenbrock function, which is widely used as a test function for optimization algorithms. 
For fixed constants $a$ and $b$, 
the function is given by
\begin{equation*} \label{eq:rosenbrock}
    f(x_1, x_2) = (a - x_1)^2 + b(x_2 - x_1^2)^2.
\end{equation*}

We take the Rosenbrock function with $a=1$ and $b=10$ as the conditional mean ($\mu_\bfx = f(x_1, x_2)$) of a Gamma distributed random variable $Y$ with shape parameter $\alpha = 100$ and scale parameter given by $\theta = \mu_\bfx / \alpha$. Assuming that $X_1 \sim \text{Uniform}(-2, 2)$ and $X_2 \sim \text{Uniform}(-1, 5)$, we generate $n = 100,000$ i.i.d. samples of $(Y, \bfX) = (Y, X_1, X_2)$ from this model. The data is split into training ($80\%$), validation ($10\%$) and test ($10\%$) data sets. The validation set is also adopted as the recalibration set. 

\begin{figure}
  \centering
  \begin{subfigure}[b]{0.25\linewidth}
    \includegraphics[width=\linewidth]{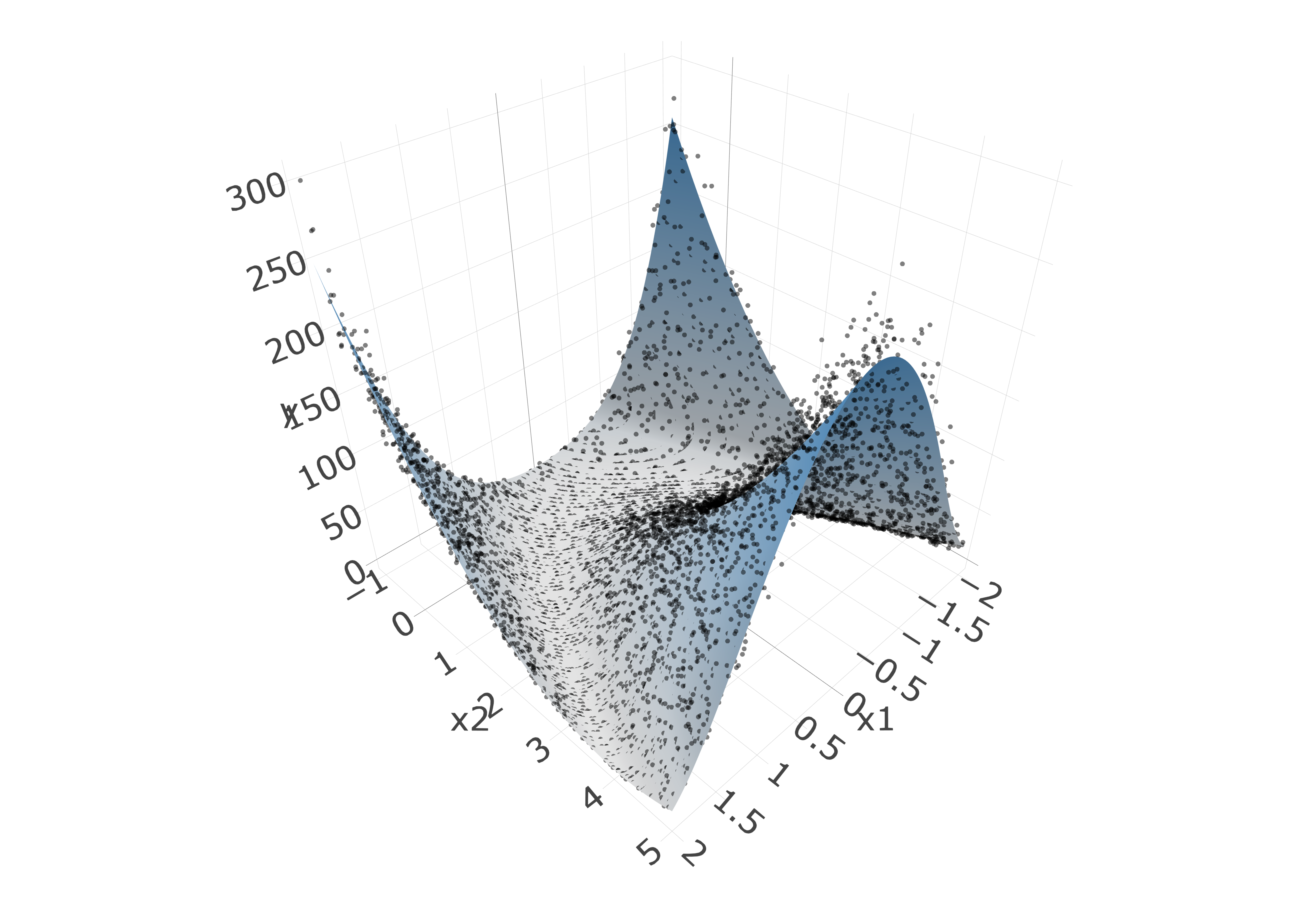}
    \caption{\footnotesize{Rosenbrock function's.}}
    \label{fig::4.2a}
  \end{subfigure}
  \vspace{.5cm}
  \begin{subfigure}[b]{0.25\linewidth}
    \includegraphics[width=\linewidth]{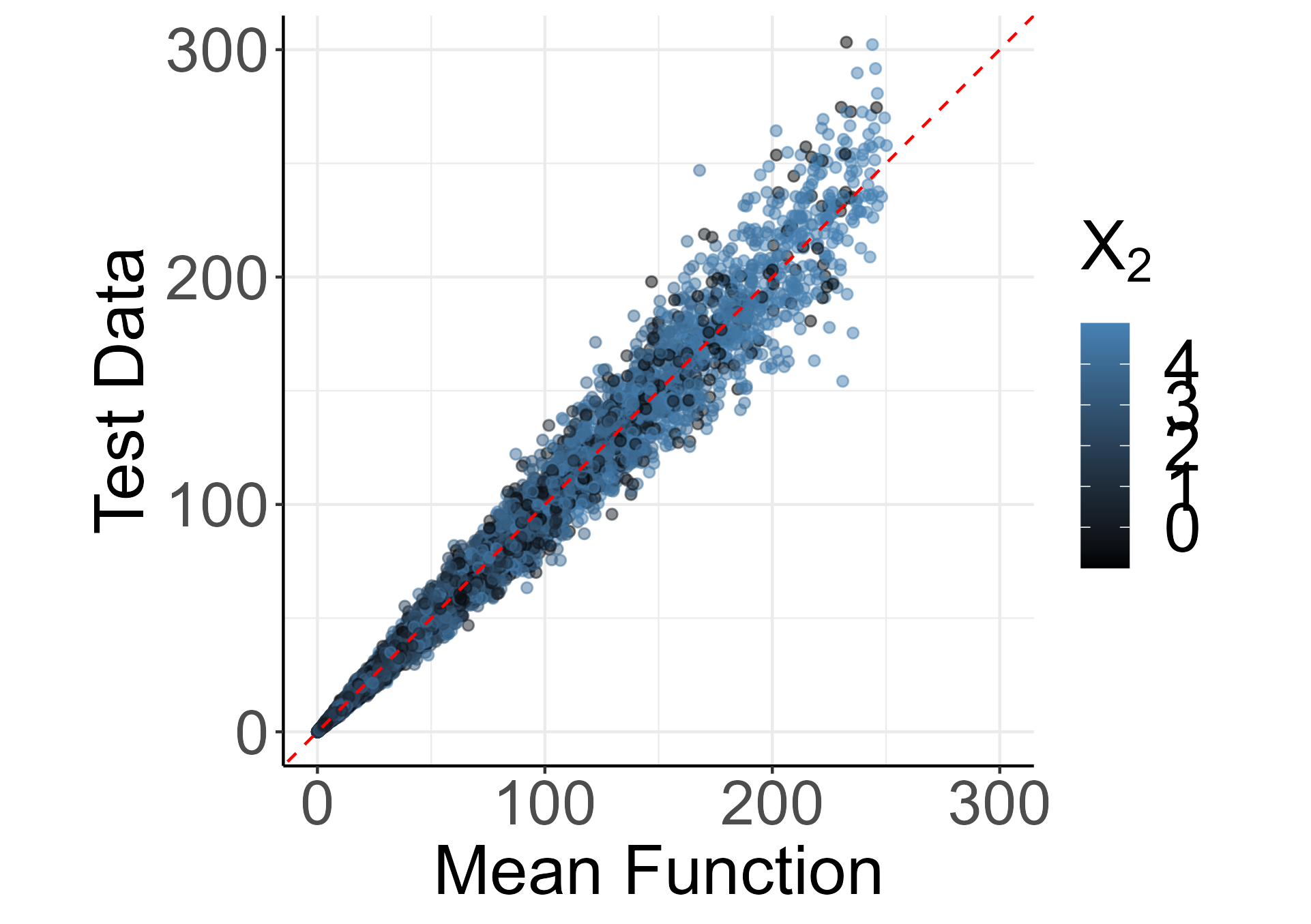}
    \caption{\footnotesize{Test data vs mean values.}}
    \label{fig::4.2b}
  \end{subfigure}
  \vspace{.5cm}
  \begin{subfigure}[b]{0.25\linewidth}
    \includegraphics[width=\linewidth]{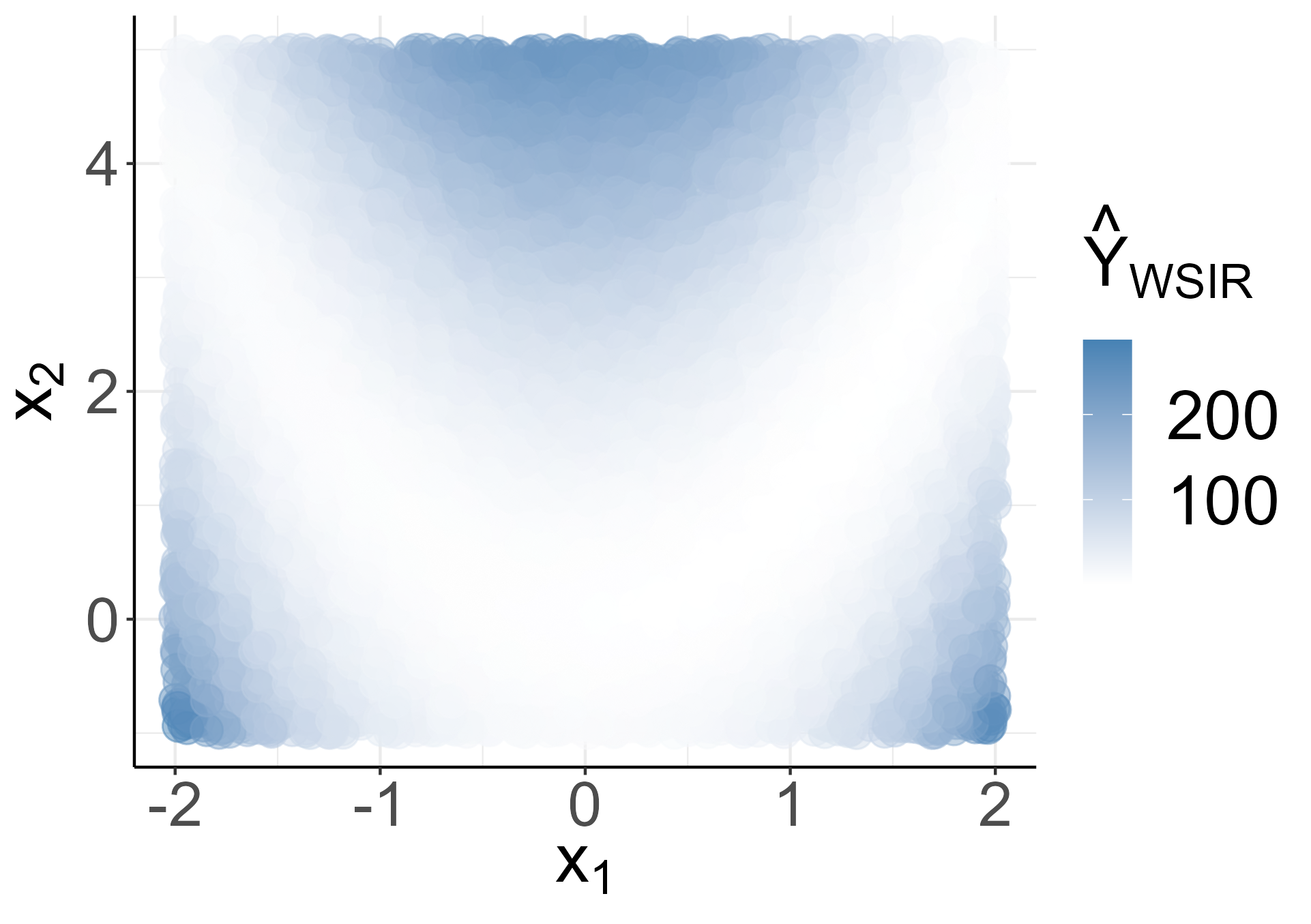}
    \caption{\footnotesize{Predictions mapping.}}
    \label{fig::4.2c}
  \end{subfigure}
  \vspace{.5cm}
  
  \begin{subfigure}[b]{0.25\linewidth}
    \includegraphics[width=\linewidth]{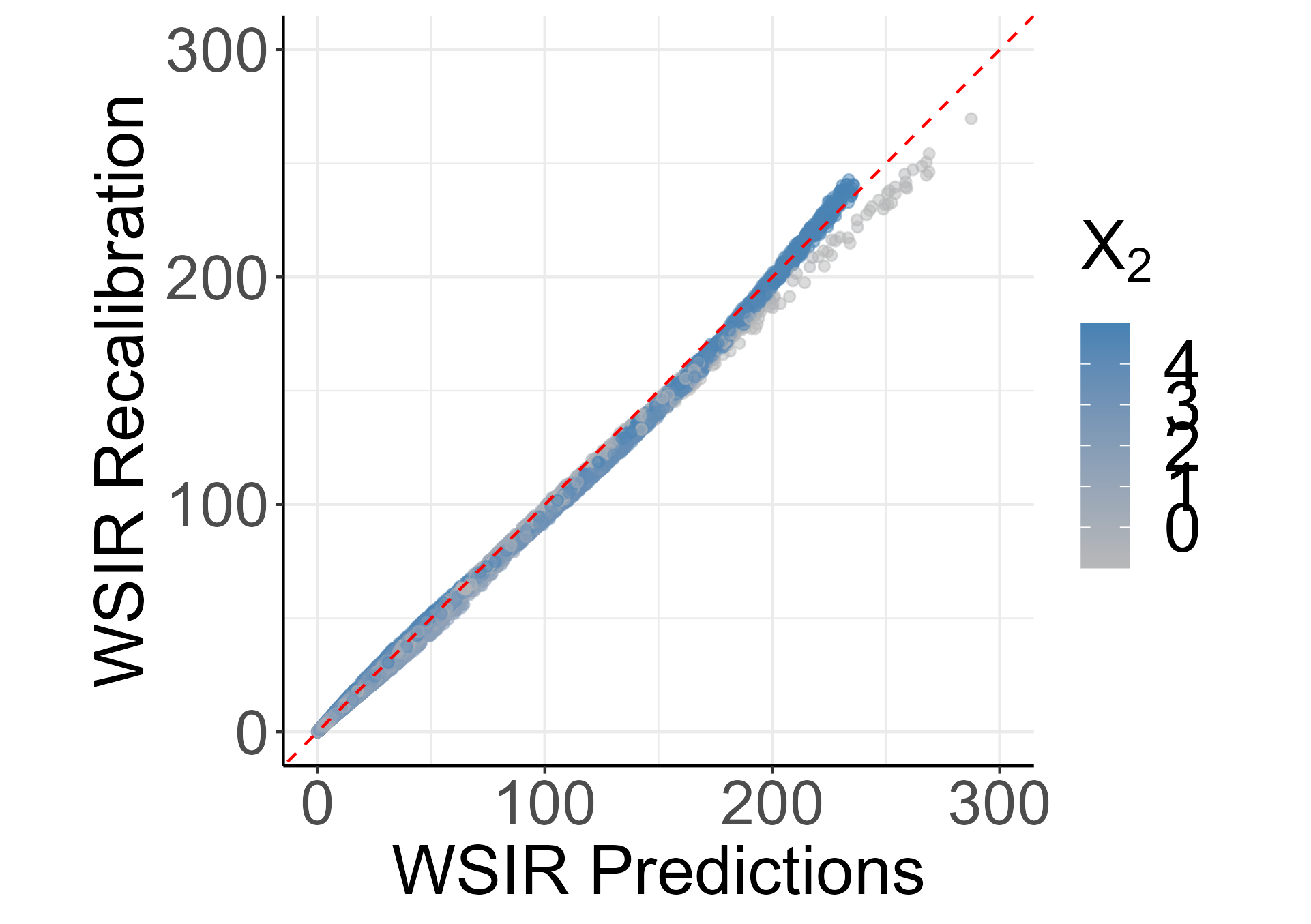}
    \caption{\footnotesize{Recalibrated predictions.}}
    \label{fig::4.2d}
  \end{subfigure}
  \vspace{.5cm}
  \begin{subfigure}[b]{0.25\linewidth}
    \includegraphics[width=\linewidth]{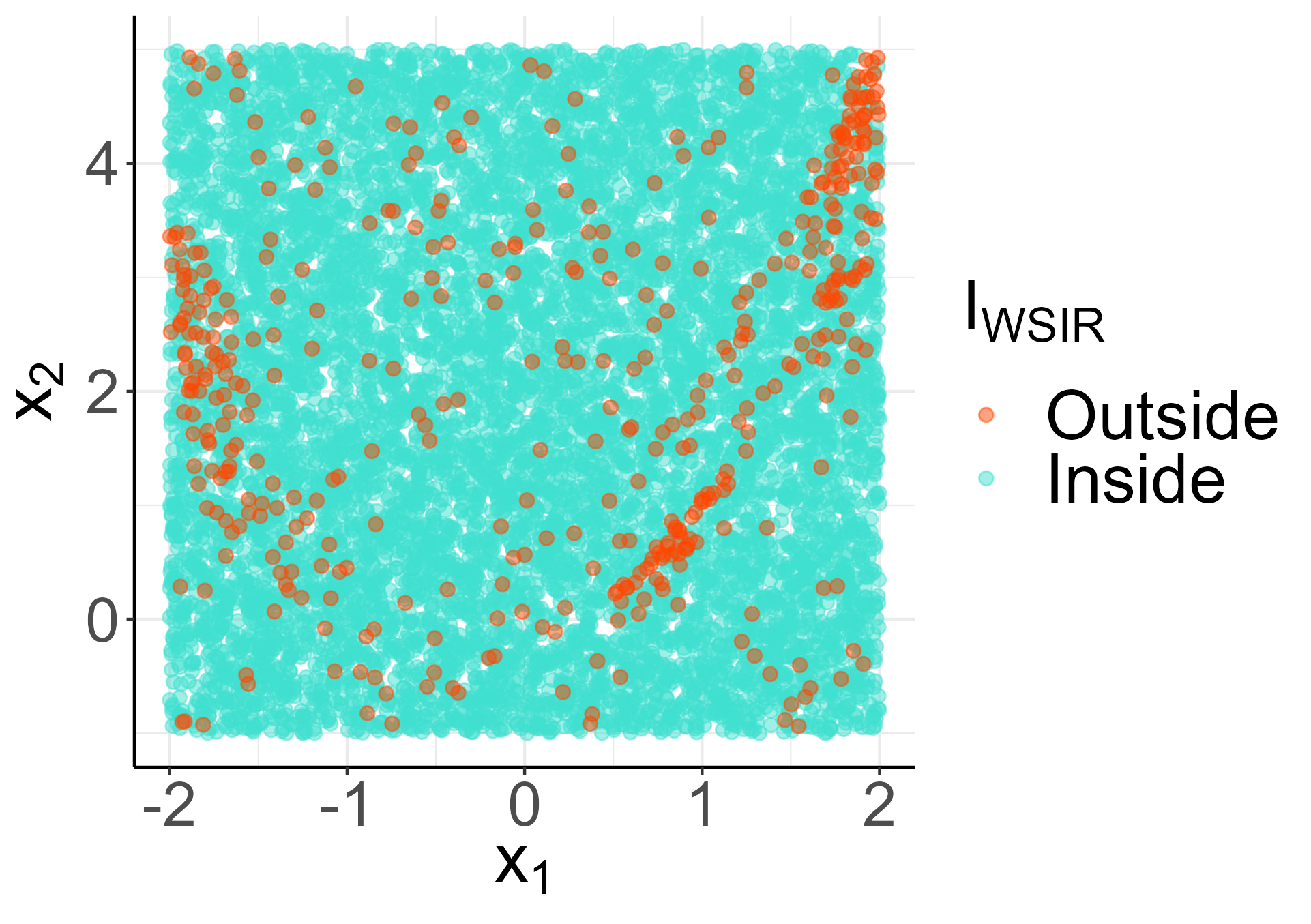}
    \caption{\footnotesize{Base model coverage.}}
    \label{fig::4.2e}
  \end{subfigure}
  \begin{subfigure}[b]{0.25\linewidth}
    \includegraphics[width=\linewidth]{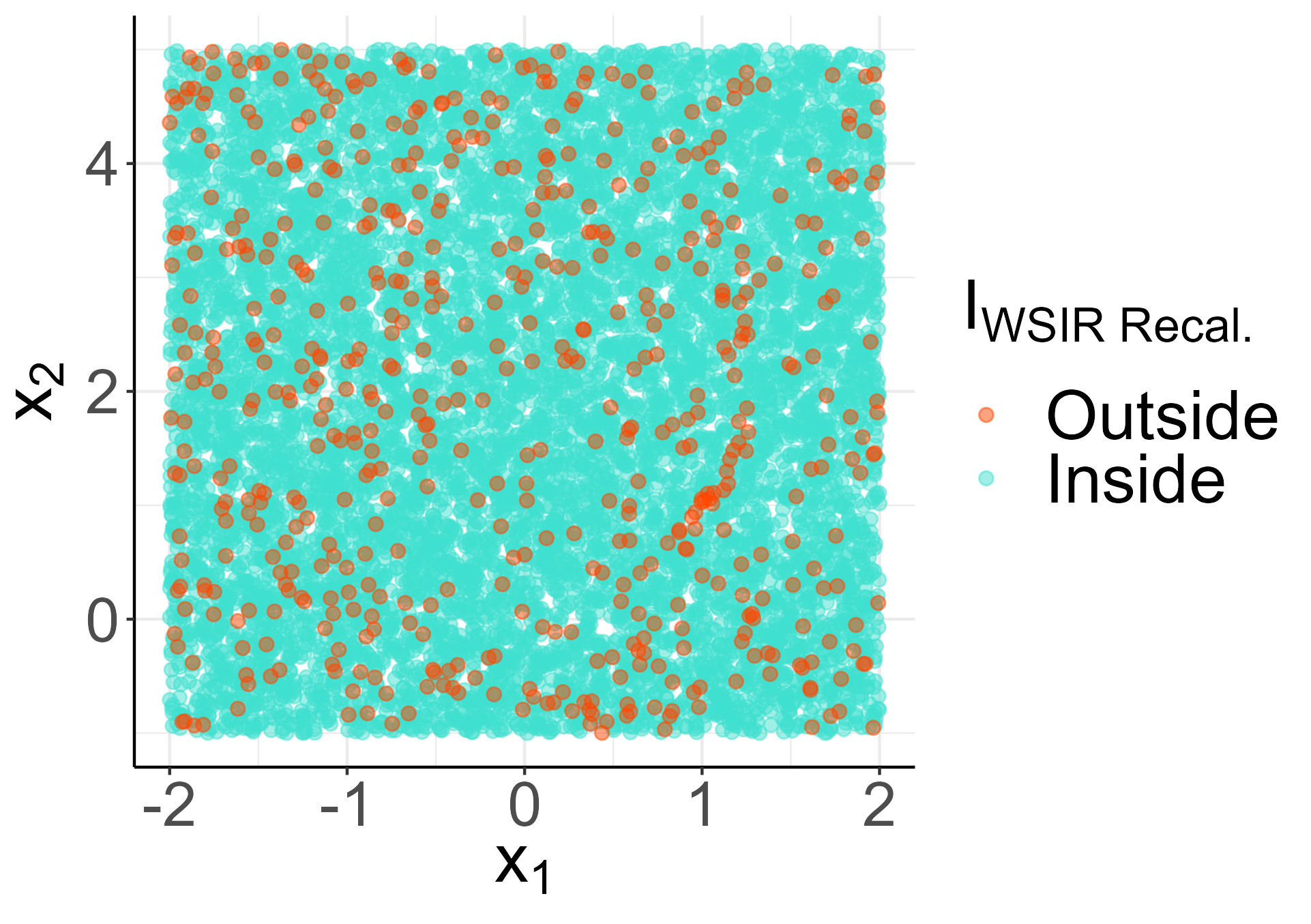}
    \caption{\footnotesize{Recalibration coverage.}}
    \label{fig::4.2f}
  \end{subfigure}
  \caption{\small (a) The non-linear Rosenbrock function's surface in three dimensions, along with test data, as a function of $\bfx$. (b) Test data as a function of the mean. (c) WSIR predictions in the covariate space. (d) WSIR predictions before and after the recalibration. (e) 95$\%$ Confidence interval misses (light red points) throughout the covariate space for the WSIR model. (f) 95$\%$ Confidence interval misses for the WSIR model with local recalibration.}
  \label{fig::4.2}
\end{figure}

Figure \ref{fig::4.2a} shows the generated test data and the mean surface in three dimensions, where it can be seen that the farther the observations are from the valley, the greater the variability around the mean. The variability of the test data can also be seen in Figure \ref{fig::4.2b}.

To estimate the mean function from the data, we fit 
a neural network model to $\log(y)$, comprising four hidden dense layers ($6400$, $6400$, $240$, $240$ neurons) with \textit{ReLU} activation function and an output linear layer with a single neuron. We also added batch normalization layers and dropout layers with probability $p = 0.5$ between every hidden layer. The network was trained for $75$ epochs with a learning rate of $0.001$, using an \textit{ADAM} optimizer, MSE loss function and batches of size $100$.

The use of dropout layers opens up two possibilities to estimate the network's uncertainty. The first approach consists of using the weight scaling inference rule (WSIR) to make a single prediction for every data point in the test set, assuming normality in the logarithmic scale of $y$. The second approach consists of using the dropout's randomly generated masks to obtain a sample of the network's predictions for every data point in the test set, following the methodology proposed by \citet{gal2016} (Monte Carlo Dropout), without parametric assumptions on the predictive distribution.

The logarithmic transformation takes the response variable from the interval $Y \in [0, \infty)$ to $\log(Y) \in (-\infty, \infty)$. Since the weight scaling inference rule averages the output of all dropout masks, in the first approach we can assume the network outputs the mean of a Normal distribution with variance equal to the validation set MSE. For every prediction, we generated a sample of size $1,000$ from the predictive distribution on the log-scale, then applied the exponential transformation, $e^{\hat{y}}$, to take the samples back to the interval $Y \in [0, \infty)$. By doing so, we obtain a Monte Carlo sample from the predictive distribution for every observation in the test set on its original scale.

The Monte Carlo Dropout (MC Dropout) approach uses the randomness intrinsic to the dropout technique to get samples of the network's predictions for each data point. At every iteration, each dropout layer generates a random mask that turns off some of the neurons according to the chosen dropout probability. We activated the dropout layers during the inference stage and generated samples of size $1,000$ for each observation in the test set (on the original scale). In both methods, the point estimates are the mean taken from the samples generated from the predictive distributions.

Since we have very high dimensionality in the network's layers weight space, we recalibrated both methods locating the nearest neighborhood of each observation in the input space. In this analysis, if we fix the number ($k$) of neighbors, the observations on the edge of the mean function's region will have a much greater neighborhood area than the ones in the center. To avoid that, for each observation in the test set, we selected the closest observations in the validation set based on a fixed maximum distance (equal to $0.5$). Due to the non-parametric nature of the samples generated by the MC Dropout method, to directly compare both method's estimates, we calculate the cumulative probabilities $p_{\text{rec}}^{(i)}$ empirically from the Monte Carlo samples obtained from each method and generated unweighted samples of the recalibrated predictive distributions according to the weights defined by the Epanechnikov kernel.

Figure \ref{fig::4.2c} shows WSIR predictions in the covariate space. One of the major effects of recalibration on both ANN models was decreased prediction values for lower values of $x_2$, previously overestimated, as illustrated with the WSIR method in \ref{fig::4.2d}. 

The performance of all methods was measured from samples obtained from their estimated predictive distributions and compared against samples taken from the true model's distribution (Table \ref{table::3.5}). While the predictions taken from the WSIR presented smaller MSE value than the ones taken from the MC Dropout, the recalibrated methods resulted in the predictions closest to the true samples by a large margin, indicating recalibrated predictions were much closer to the original test data on average. In regards to $95\%$ prediction interval
estimation, the closest to nominal coverage and largest interval score were attained by the MC Dropout method, with
the latter suggesting this method generated too wide prediction intervals. The recalibrated methods had coverage and score metrics very close to 
those of the true model. 

\begin{table}
\footnotesize
\centering
\begin{tabular}{lccc}
\hline
Model                    & MSE     & Coverage (\%) & sMIS   \\
\hline
WSIR ANN                 & 71.25 & 95.5   & 0.569 \\
MC Dropout ANN           & 75.06 & 99.0   & 0.874 \\
Recalibrated WSIR ANN    & 60.52 & 94.3   & 0.500 \\
Recalibrated MC Dropout  & \textbf{60.44} & \textbf{94.7} & \textbf{0.499} \\
True model               & 59.64 & 94.6   & 0.477 \\
\hline
\end{tabular}
\caption{\small Performance comparison between all methods.}
\label{table::3.5}
\end{table}

The coverage of the WSIR methods (with and without recalibration) is shown in Figures \ref{fig::4.2e} and \ref{fig::4.2f}, where light red points represent observations not captured by the prediction intervals. It can be seen that ANN predictions, despite getting close to the nominal coverage level presented a very clear pattern, missing observations in specific areas. On the other hand, recalibration appears to have largely reduced local biases. That is, after the recalibration, the coverage showed a much more evenly spread pattern throughout the covariate space, an indication that these models are well-calibrated.

\subsection{Simulation Study}
\label{sec:simulation}

In this analysis, we run a simulation study to investigate and quantify the effect of our recalibration procedure on neural network models under various conditions. We revisit the analysis in \citet{tran2020} and consider the highly nonlinear model given by
\begin{equation}\label{simmodel}
y = 5 + 10x_1 + \frac{10}{x_2^2 + 1} + 5x_3x_4 + 2x_4  + 5x_4^2 + 5x_5 + 2x_6 + \frac{10}{x_7^2 + 1} + 5x_8x_9 + 5x_9^2 + 5x_{10} + \delta,
\end{equation}
where $\delta \sim N(0, 1)$. We generate the variables $x_1, \dots, x_{20}$, from which $x_{11}, \dots, x_{20}$ are non-informative to the model, from a multivariate normal distribution with mean vector $\mathbf{0} = (0)_i$, $i = 1, \dots, 20$, and covariance matrix $(0.5^{|i - j|})_{i, j}$, $i = 1, \dots, 20, j = 1, \dots, 20$. Data from this process are simulated in sets of sizes $N = 10^{3}$, $10^{4}$, $10^{5}$ and $10^{6}$. Then, in each scenario, we randomly split the data into a training set, a validation set and a test set using $80\%$, $10\%$ and $10\%$, respectively. We propose to fit and recalibrate two neural network models as well as compare recalibration performance with a K-Nearest Neighbor (KNN) regression model, due to methodological similarities, and two well-known post-hoc recalibration methods, isotonic regression recalibration \citep{kuleshov2018} and density estimation recalibration \citep{kuleshov2022}. The simulation study is run multiple times, each with a different seed.

Because this section compares the models' performances for each configuration under study, all recalibration models and the KNN regression model are fit with the validation set and evaluated with the test set, it not being necessary to optimize KNN regression parameter values.

Apart from the number of neurons, the two neural networks considered are identical in architecture, both being composed of a $20$ neurons input layer, followed by four hidden layers with \textit{ReLU} activation function, then a linear output layer with a single neuron. The "small" network's hidden layers are all composed of $5$ neurons, while the "big" network's first three hidden layers have $200$ neurons and its last hidden layer has $5$ neurons. The neural networks are trained with the \textit{Adam} optimizer and mean squared error loss function with the early stop callback. Due to computational and time constraints, both neural network model's learning rates were optimized in advance for every scenario considered in the simulation, with values ranging from $10^{-4}$ to $10^{-2}$.

The neural network models are recalibrated assuming normally distributed predictive distributions and considering Epanechnikov's kernel function to weight the recalibrated samples. To evaluate the effects of recalibration, both networks are recalibrated on the input layer, the fourth layer, the fifth layer and the output layer.

The networks are recalibrated with the $100$, $500$ and $1000$ nearest neighbors. The nearest neighbor search is conducted with three degrees of approximation: $\epsilon = 0$ (exact search), $\epsilon = 0.5$ and $\epsilon = 1$, where $\epsilon$ is the KNN search approximation parameter. 

For the density estimation models, we replicate the same quantile neural network architecture in the experiments described by \citet{kuleshov2022}. The models are trained with $80\%$ of the original validation set and validated with the last $20\%$. Predictions and metrics are taken from the test set, as with the other models used in this simulation.

\begin{figure}
  \centering
  
  \begin{subfigure}[b]{0.4\linewidth}
    \includegraphics[width=\linewidth]{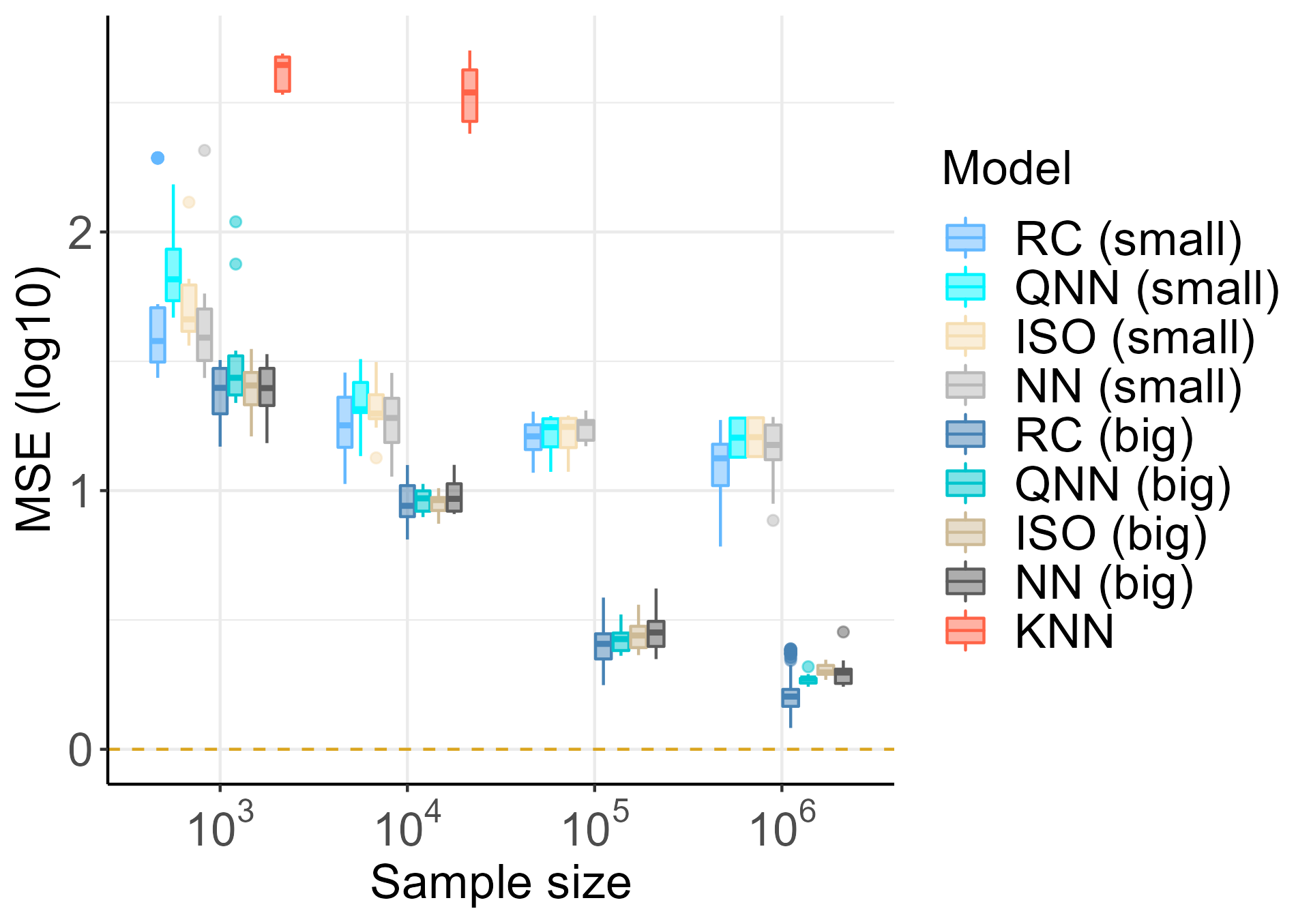}
    \caption{\footnotesize{MSE by data size.}}
    \label{fig::4.3.a}
  \end{subfigure}
  \vspace{.5cm}
      \begin{subfigure}[b]{0.4\linewidth}
    \includegraphics[width=\linewidth]{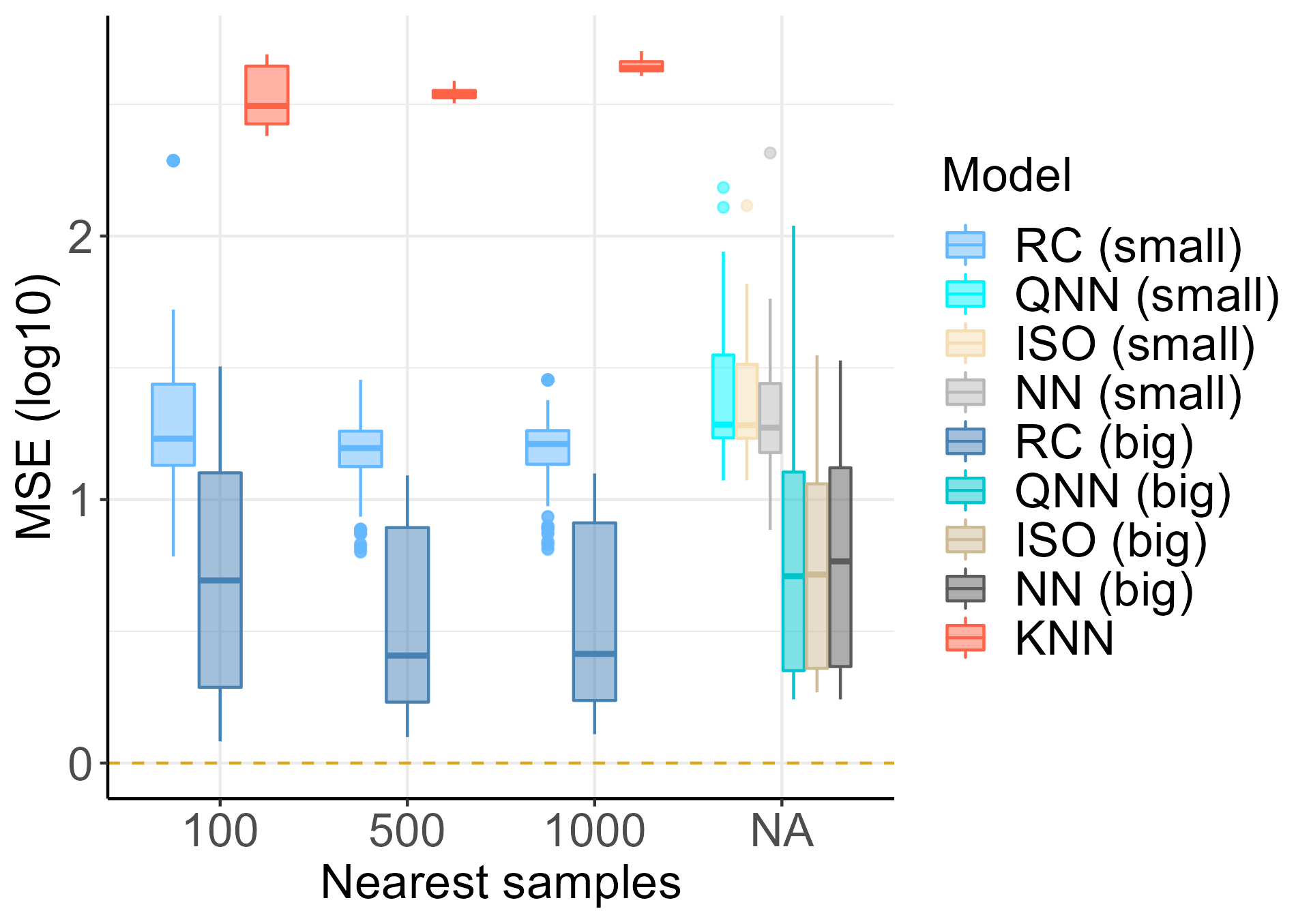}
    \caption{\footnotesize{MSE by nearest samples.}}
    \label{fig::4.3.b}
  \end{subfigure}
  \vspace{.5cm}
  
  \begin{subfigure}[b]{0.4\linewidth}
    \includegraphics[width=\linewidth]{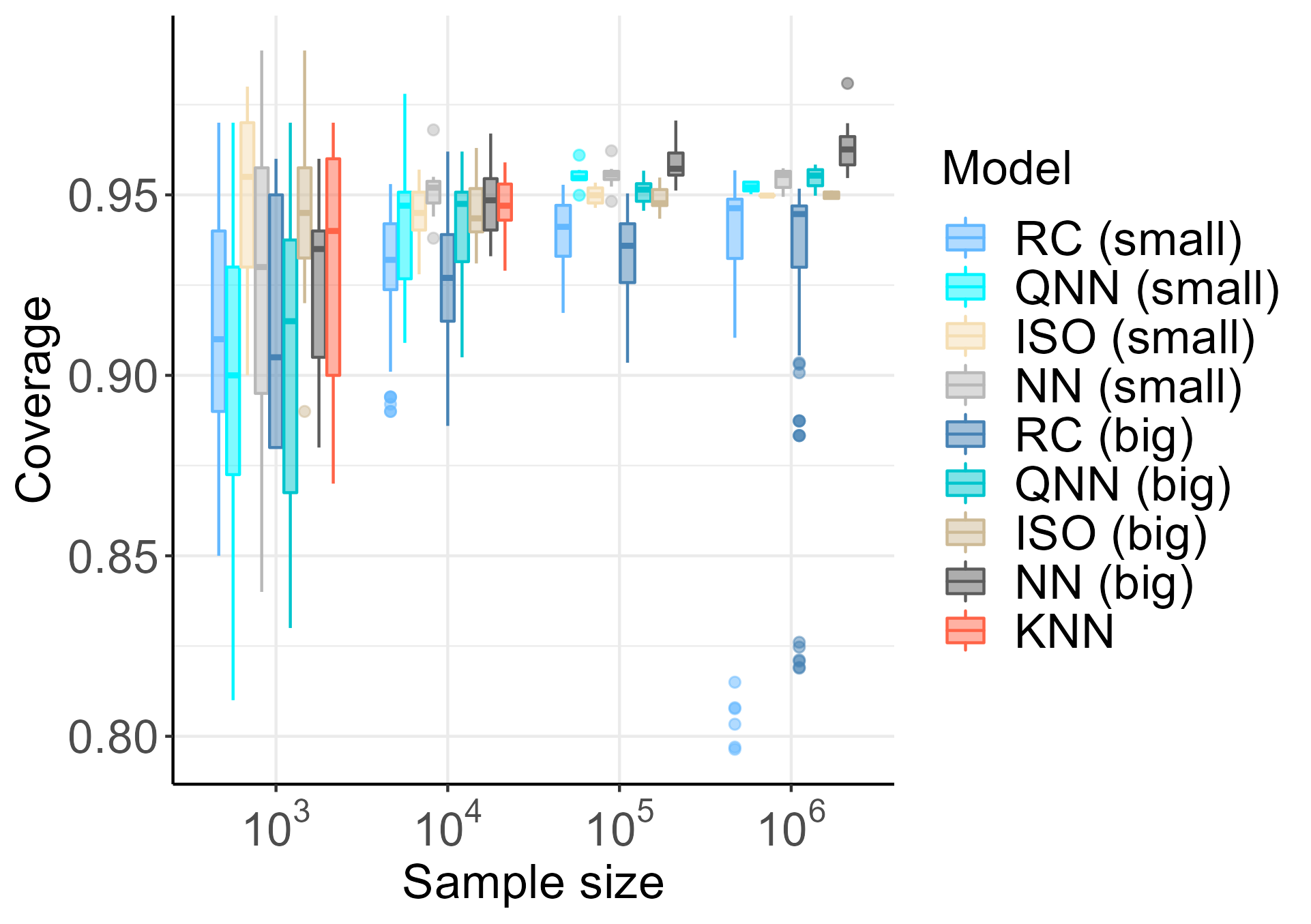}
    \caption{\footnotesize{Coverage by data size.}}
    \label{fig::4.3.c}
  \end{subfigure}
  \vspace{.5cm}
    \begin{subfigure}[b]{0.4\linewidth}
    \includegraphics[width=\linewidth]{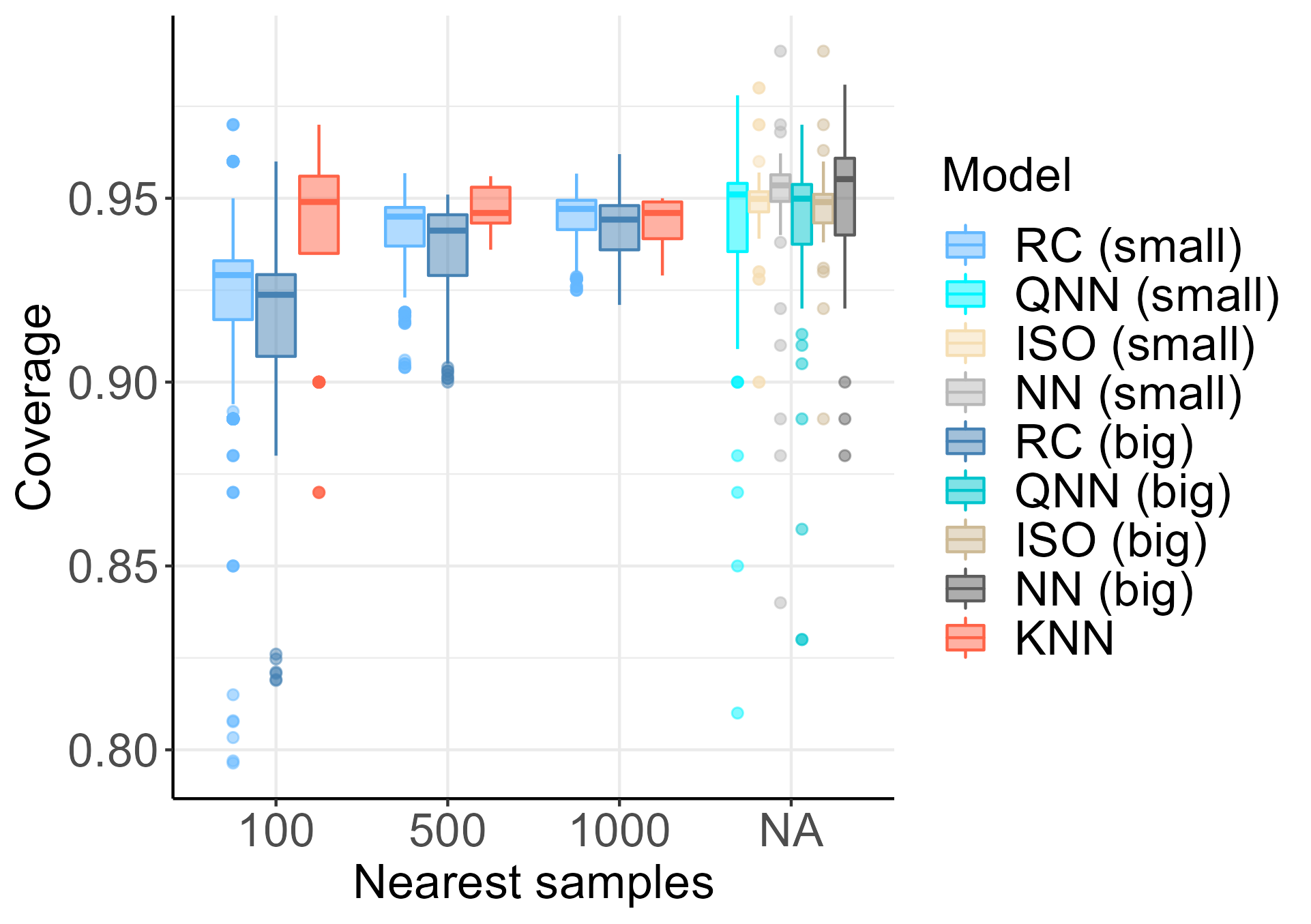}
    \caption{\footnotesize{Coverage by nearest samples.}}
    \label{fig::4.3.d}
  \end{subfigure}
  \vspace{.5cm}
  
    \begin{subfigure}[b]{0.4\linewidth}
    \includegraphics[width=\linewidth]{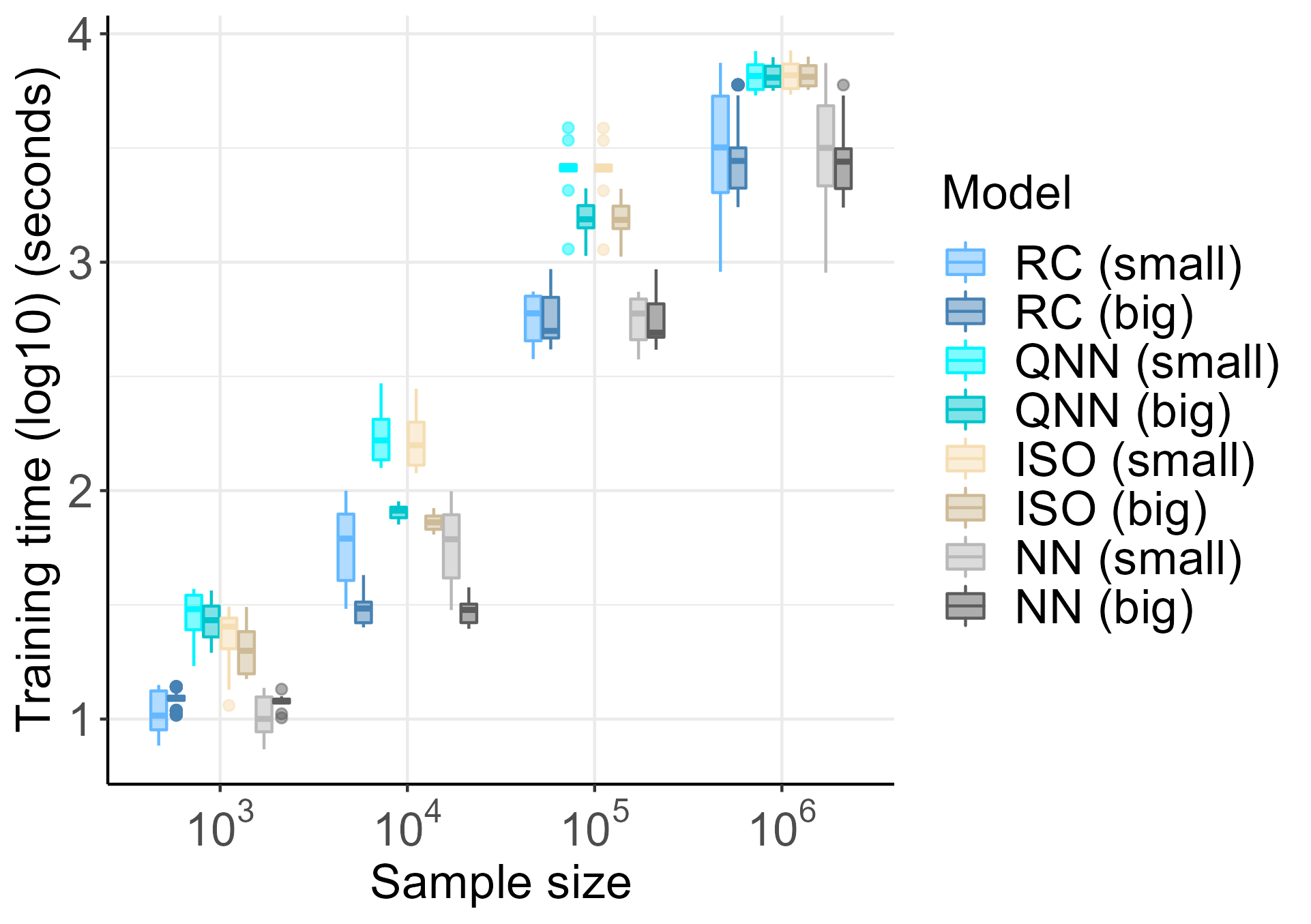}
    \caption{\footnotesize{Training time by data size.}}
    \label{fig::4.3.e}
  \end{subfigure}
    \begin{subfigure}[b]{0.4\linewidth}
    \includegraphics[width=\linewidth]{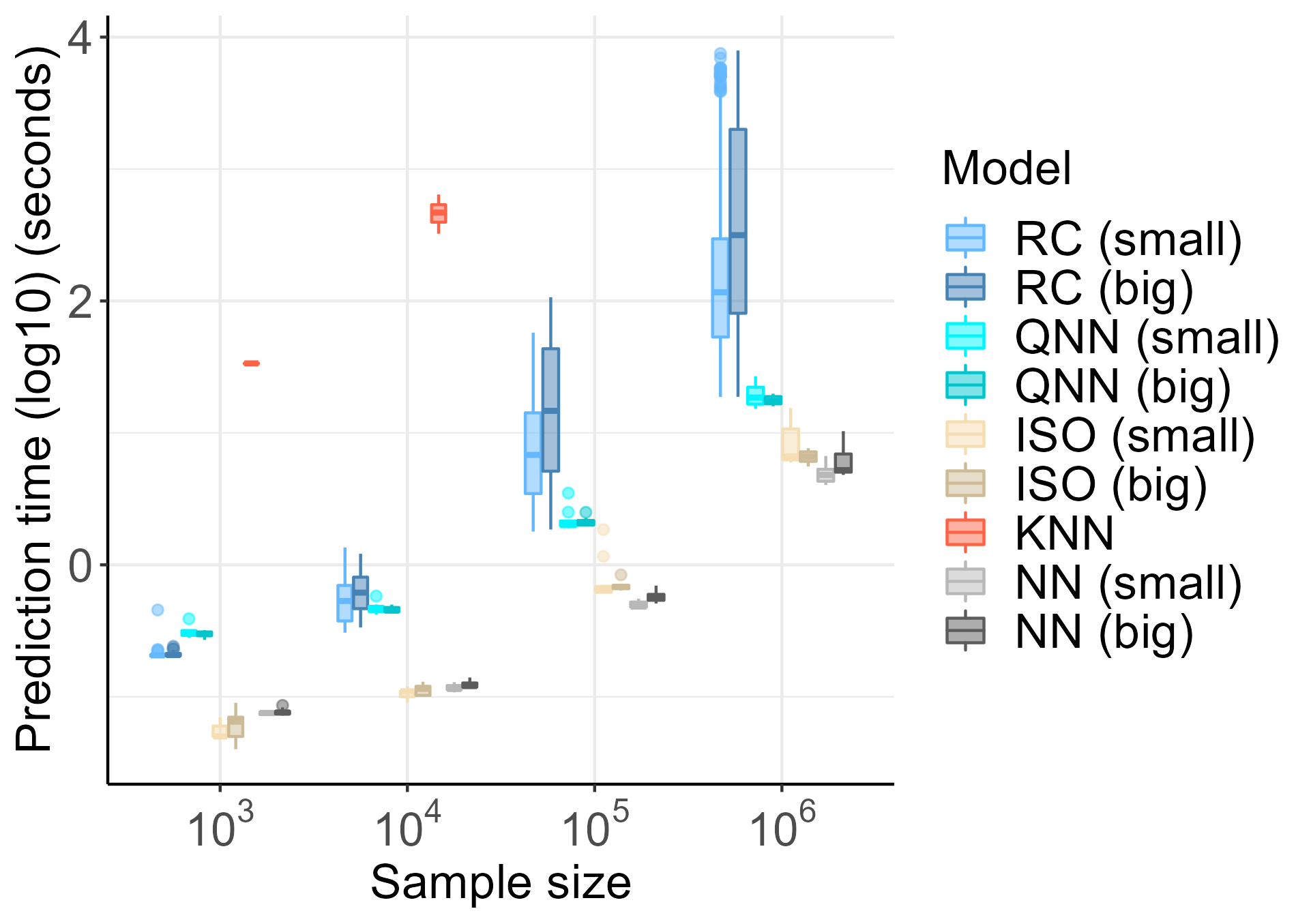}
    \caption{\footnotesize{Prediction time by data size.}}
    \label{fig::4.3.f}
  \end{subfigure}
  
    \caption{\small (a) and (b) show boxplots of the observed MSE for each sample and nearest neighbours samples sizes, respectively, for neural networks (NN), isotonic regression (ISO), density estimation (QNN) and our novel recalibration (RC) method. (c) and (d) present $95\%$ interval coverage for various scenarios. (e) and (f) exhibit 
 training and prediction times, respectively, according to the sample size.}
  \label{fig::4.3}
\end{figure}

An algorithm directly comparable to the proposed recalibration method is the KNN regression, which we also apply considering the Epanechnikov's kernel function, the same KNN approximate search algorithm and the same set of values of the $k$ (nearest samples) and $\epsilon$ parameters, for direct comparison. Due to memory and computational time constraints, all recalibrated models and KNN regression were fit in batches according to test set size, KNN regression being applied only to the first two scenarios of $N = 10^3$ and $N = 10^4$. In the last two scenarios, KNN regression computational cost was simply too high to be carried over, taking alone up to two-thirds of the total simulation time, on average. As shown in Figure \ref{fig::4.3} though, it is clear that, while KNN regression is comparable to the recalibration in terms of methodology, it does not compare in terms of computational efficiency and model performance.

Figure \ref{fig::4.3.a} shows the models' MSE as sample size increases, with the true model MSE (equal to 1) represented by the golden dashed line. Overall, recalibration improved MSE over the network models and other recalibration methods, with larger impact on the smaller neural network, benefiting the most when there is plenty of data. It can be observed that even with more data provided, the smaller neural network's low capacity prevents it from improving model MSE further. Even in this case, recalibration was able to improve performance, on average. In Figure \ref{fig::4.3.b}, increasing the nearest samples size improved the biggest recalibrated model's MSE on average with diminishing effects as nearest samples size increases, meaning that increasing the proportion of nearest data will not add much information to the model predictions when there is already enough information, in terms of model capacity. The target layer and the approximation level also did not influence much the prediction accuracy (figure not shown).
Figures \ref{fig::4.3.c} and \ref{fig::4.3.d} show the observed coverage of $95\%$ confidence intervals for different settings. Unsurprisingly, all models become more reliable as the sample size increases, and our recalibration method (RC), in particular, becomes more consistent for higher values of the parameter $k$.

Figure \ref{fig::4.3.e} shows that the proposed recalibration algorithm does not significantly affect model training time as the sample sizes increase. However, Figure \ref{fig::4.3.f} confirms that our proposal significantly affects prediction time. This is partially mitigated by the search level of approximation, with the exact search taking longer than a higher approximation level (figure not show). Recalibration prediction memory size grows in $O(n)$, proportional to $n(2k + h + d)$, where $n$ is the number of test observations, $k$ is the number of nearest samples, $h$ is the number of neurons in the target layer and $d$ is the size of the network's output. 

\subsection{Diamonds Price Prediction}
\label{sec:diamonds}

We explore the data from $53,940$ diamonds from the dataset \textit{"diamonds"} available in the \textit{ggplot2} R package \citep{hadley2016}. This data set lists several diamond characteristics including price in US dollars, carat weight, cut quality, colour, clarity, length (in mm), width (mm), depth (mm), and table width (width of the top of the diamond relative to the widest point).

Interest is in modelling diamond prices conditional on their physical attributes. 
We specify the response variable as Gamma distributed, $Y|\bfX \sim \text{Gamma}(\alpha, \mu/\alpha)$, with $\mu = \mathbb{E}(Y|\bfX)$ the conditional mean and $\alpha$ the shape parameter. 
The data were randomly assigned to the training set (70\%), validation/recalibration set (20\%), and test set (10\%). We consider two models -- a generalized linear Gamma model (GLM) with logarithmic link function, and a neural network model with negative Gamma log-likelihood loss function. 

The architecture of the neural network model was selected by minimizing the loss function in the validation set within a predefined search space. The final network architecture is composed of an input layer, a hidden layer with $100$ neurons and \textit{ReLu} activation function, two hidden layers with $5$ neurons and \textit{ReLu} activation function, and a forked output with two exponential layers, one for estimating the mean, $\mu$, and the other with zero-constrained weights for estimating the shape parameter, $\alpha$, which is the same for all observations. 

In this application, we recalibrate both the Generalized Linear Model and the Neural Network by employing the three distinct methods considered in the simulation study presented in Section 4.2: Isotonic Regression Recalibration \citep{kuleshov2018}, Density Estimation Recalibration \citep{kuleshov2022}, and our Quantile Recalibration approach. Inspired by findings from the simulation study and considering our low-dimensional feature vector, we execute the Quantile Recalibration locally, directly within the input space ($l=1$), using $k = 1,000$ nearest neighbors.

Table \ref{table::4.4} illustrates the impact of recalibration on prediction accuracy, measured by the Root-mean-square deviation, and the observed coverage for confidence intervals at $90\%$, $95\%$, and $99\%$. Quantile Recalibration stands out as the sole method that effectively decreased the Root Mean Square Error (RMSE) of the Neural Network. Additionally, both Isotonic Regression Recalibration and Quantile Recalibration demonstrated remarkable performance in achieving observed coverage close to their respective nominal levels.

\begin{table}
\footnotesize
\centering
\begin{tabular}{lcccc}
\hline
\multirow{2}{*}{Model } & \multirow{2}{*}{RMSE} & \multicolumn{3}{c}{Observed Coverage} \\
                      &                       & 90\%     & 95\%    & 99\%    \\
\hline
GLM                       & 798.9 & 0.955 & 0.982 & 0.997 \\
Isotonic GLM RC           & 789.3 & 0.894 & \textbf{0.949} & \textbf{0.991} \\
Density Estimation GLM RC & 796.0 & 0.956 & 0.982 & 0.997 \\
Quantile GLM RC           & \textbf{751.2} & \textbf{0.898} & 0.946 & 0.986 \\
\hline
NN                        & 557.8 & 0.933 & 0.966 & 0.987 \\
Isotonic NN RC            & 557.8 & \textbf{0.898} & \textbf{0.947} & \textbf{0.988} \\
Density Estimation NN RC  & 558.0 & 0.934 & 0.965 & 0.987 \\
Quantile NN RC            & \textbf{542.8} & 0.896 & 0.944 & 0.986 \\
\hline
\end{tabular}
\caption{\small Performance comparison between base models and recalibrated (RC) models showing the overall performance improvement and the observed coverage for confidence intervals at $90\%$, $95\%$, and $99\%$.}
\label{table::4.4}
\end{table}

\begin{figure}
  \centering
  
  \begin{subfigure}[b]{0.4\linewidth}
    \includegraphics[width=\linewidth]{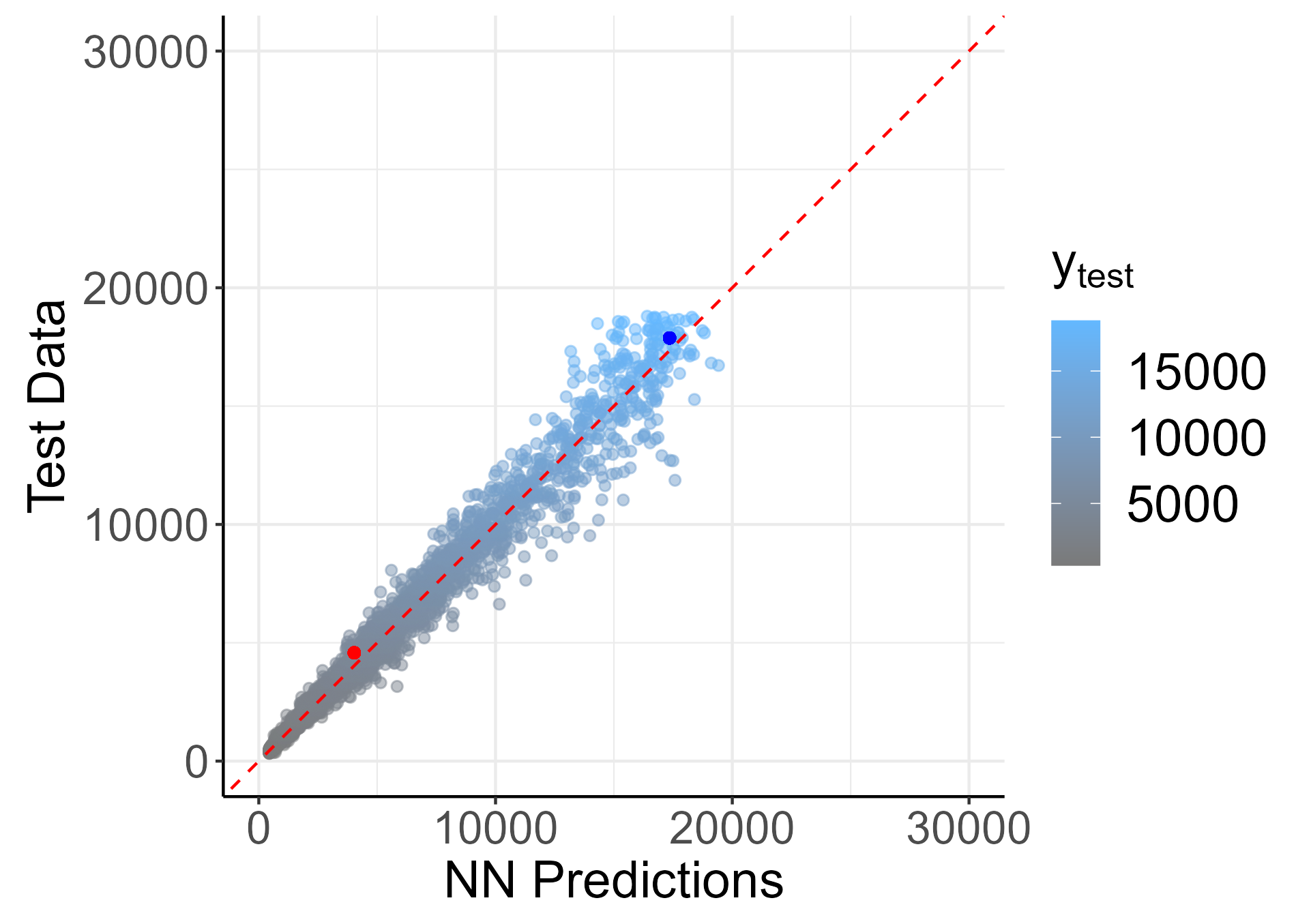}
    \caption{\footnotesize{NN predictions.}}
    \label{fig::4.4a}
  \end{subfigure}
  \vspace{.5cm}
  \begin{subfigure}[b]{0.4\linewidth}
    \includegraphics[width=\linewidth]{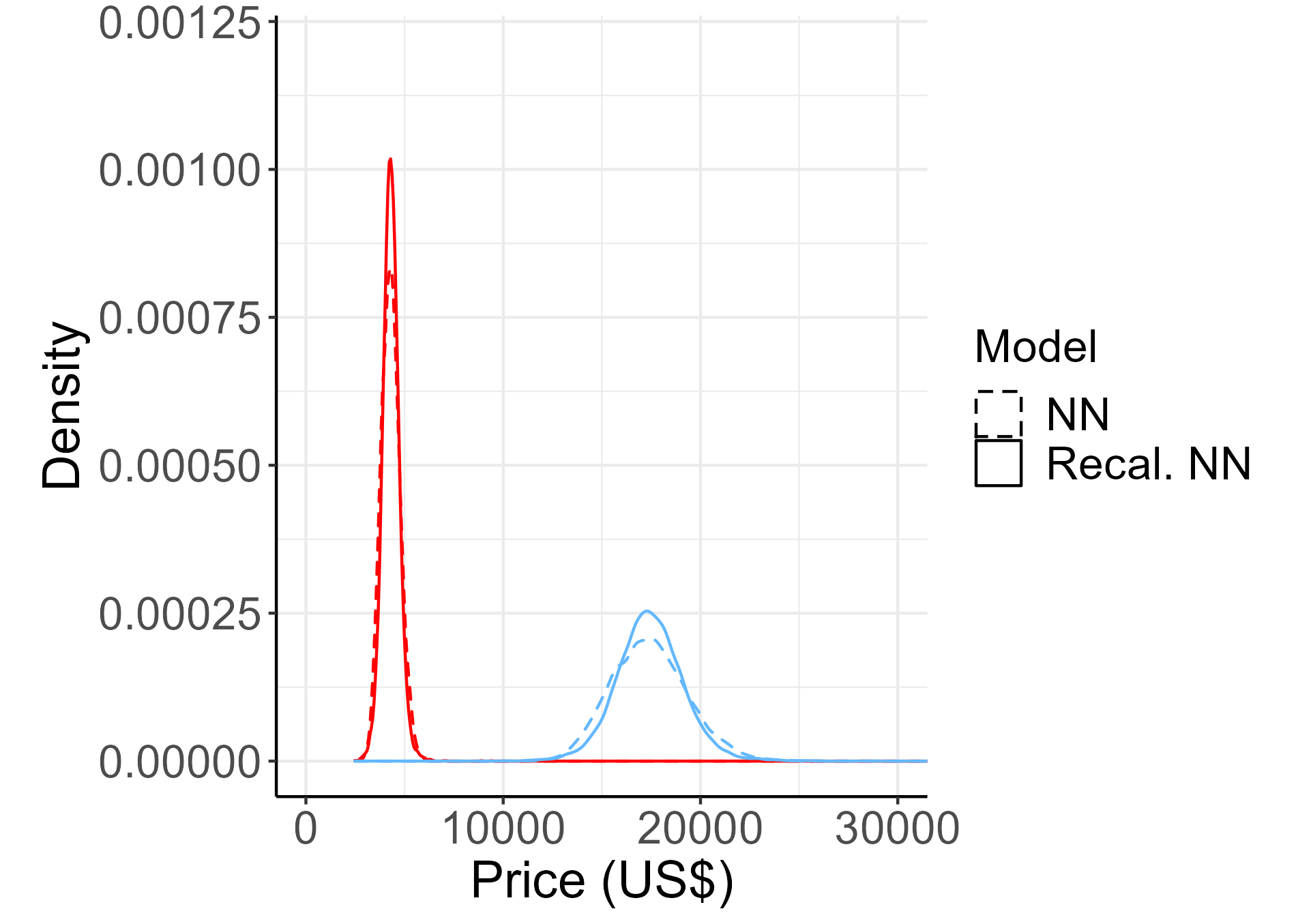}
    \caption{\footnotesize{$y^{(0)}_{test}$ and $y^{(1)}_{test}$ predictive distributions.}}
    \label{fig::4.4b}
   \end{subfigure}
   \vspace{.5cm}

  \begin{subfigure}[b]{0.4\linewidth}
    \includegraphics[width=\linewidth]{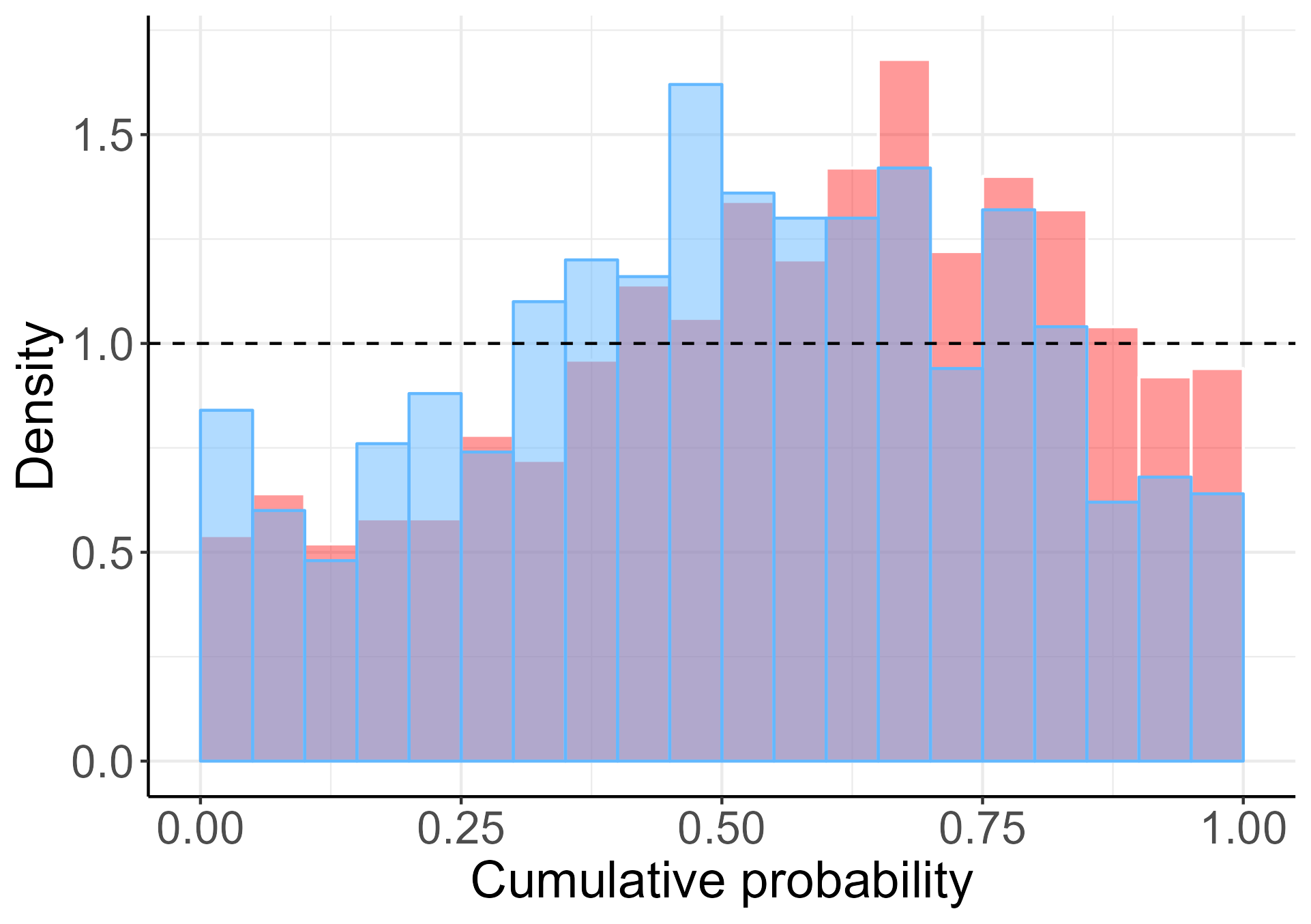}
    \caption{\footnotesize{NN local cumulative probabilities.}}
    \label{fig::4.4c}
  \end{subfigure}
  \vspace{.5cm}
  \begin{subfigure}[b]{0.4\linewidth}
    \includegraphics[width=\linewidth]{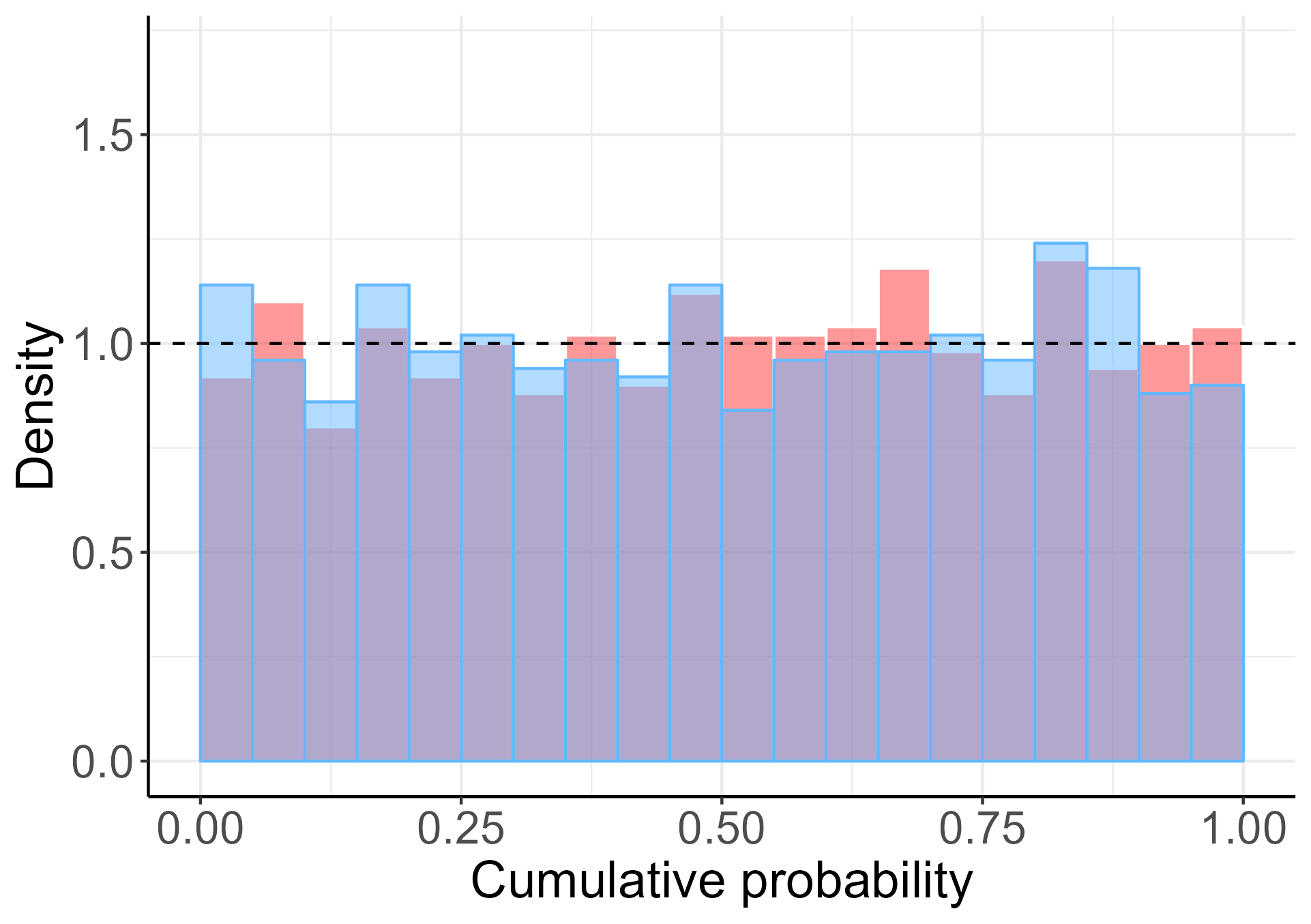}
    \caption{\footnotesize{Recalibrated local cumulative probabilities.}}
    \label{fig::4.4d}
  \end{subfigure}
  
  \caption{\small (a) Predictions variability of the base model in relation to the true data, with highlighted points $y^{(0)}_{test}$ (red) and $y^{(1)}_{test}$ (blue) for reference. (b) Estimated predictive distributions before and after recalibration for $y^{(0)}_{test}$ and $y^{(1)}_{test}$. (c) and (d) Local cumulative probabilities in the neighborhood of the highlighted points for the base and recalibrated models, respectively.}
  \label{fig::4.4}
\end{figure}

Figure \ref{fig::4.4} further explores the impact of our recalibration strategy on the base Neural Network model across different regions of the dependent variable space. Specifically, Figures \ref{fig::4.4a} compares predictions made by the base model with observed diamond prices in the test set. There is an increased prediction dispersion for higher values of $Y$, underscoring the necessity for a heteroscedastic model. To illustrate local patterns, two points, $y^{(0)}_{\text{test}} = 4,113$ (red) and $y^{(1)}_{\text{test}} = 18,026$ (blue), are highlighted. Figure \ref{fig::4.4b} overlays the predictive distributions of the base and recalibrated models for each highlighted point, showcasing that recalibration has not only decreased the variance but also adjusted the distribution mean and shape in each case. 

The cumulative probability histograms in Figure \ref{fig::4.4c} reveal the deviation from uniformity for the base model in the proximity of the highlighted points. In contrast, the histogram of the recalibrated model (Figure \ref{fig::4.4d}) exhibits a more uniform distribution, emphasizing that the recalibrated model provides a more accurate probabilistic representation of the conditional distributions.

\section{Conclusion}
\label{sec:conclusion}

This work introduces a novel method for recalibrating neural networks based on the observed cumulative probabilities of their predictive distributions. Unlike previous approaches that only consider the network's outputs, this method addresses region-specific bias patterns by recalibrating the networks locally within a space learned by the neural network itself. Two different methods are considered when creating the initial probabilistic neural networks. The first involves assuming a probabilistic model through the selection of an appropriate loss function, while the second employs an empirical predictive distribution obtained via Monte Carlo Dropout. 

The results of simulations using simple examples demonstrate the positive impact of the proposed method on various performance metrics. Recalibrated models exhibit improved Mean Squared Error, confidence intervals coverage, and interval scores compared to uncalibrated models.  These improvements indicate better prediction accuracy and closer approximation to the true distribution. Furthermore, the recalibrated models demonstrate evenly distributed interval coverage while reducing interval width, effectively correcting local prediction bias and improving variance estimation. 

Through a simulation study we explored the effects of different recalibration parameter configurations and compared the performance of our procedure with alternative recalibration approaches \citep{kuleshov2018, kuleshov2022}.
The study revealed that training time is primarily influenced by the number of observations in the training set, while prediction time increases with nearest samples size, $k$, and decreases with target layer dimensionality (number of neurons in the recalibration layer). Importantly, recalibration consistently yields superior MSE and interval metrics. Moreover, recalibrated neural networks prove to be more efficient than the simple application of KNN regression under the same conditions. 

When tested against real data, recalibration demonstrates positive effects on prediction precision and confidence interval metrics. Sections \ref{sec:gaussian}, \ref{sec:gamma} and \ref{sec:diamonds} demonstrate that recalibration can offer advantages even when the observed coverage closely matches the nominal level. This is particularly pertinent because the model may be locally uncalibrated, potentially leading to a misleading characterization of the underlying system. Additionally, it's important to note that recalibration methods are not explicitly designed to minimize Mean Squared Error (MSE) or another specific metric. Instead, their primary goal is to offer a more comprehensive and accurate probabilistic description of the data-generative process.

Although our approach successfully improves the quality of predictive distributions, it is important to acknowledge that the prediction time can be a potential drawback. As a result, this method may not be suitable for scenarios where fast predictions are of utmost importance.

\bibliographystyle{JASA}
\bibliography{Bibliography.bib}

\end{document}